\documentclass[12pt]{article} 
\usepackage{fancybox}
\usepackage{graphicx}
\usepackage{epsf}
\usepackage{amsmath,amssymb}
\newcommand{\lsim}{\raisebox{-0.13cm}{~\shortstack{$<$
 \\[-0.07cm] $\sim$}}~}
\newcommand{\gsim}{\raisebox{-0.13cm}{~\shortstack{$>$
 \\[-0.07cm] $\sim$}}~}
 
\begin{document} 
 
%
\thispagestyle{empty}
\setcounter{page}{0}
\vspace{10mm}
\begin{center}
{\Large{\bf Testing supersymmetry at the LHC through \\[1.01ex]
         gluon-fusion production of a slepton pair }}\\[10mm]
{\large 
 F.~Borzumati$^{1}$ and K.~Hagiwara$^{2}$} 
\end{center}
\begin{center}
$^1$ {\it Department of Physics, National Taiwan University,
          Taipei 10617, Taiwan}
\\[1mm] 
$^2${\it 
 KEK Theory Division, and Sokendai, Tsukuba 305-0801, Japan}
\end{center}
\vskip 2cm

\begin{abstract}
Renormalizable quartic couplings among new particles are typical of
 supersymmetric models.
Their detection could provide a test for supersymmetry, discriminating
 it from other extensions of the Standard Model. 
Quartic couplings among squarks and sleptons, together with the SU(3)
 gauge couplings for squarks, allow a new realization of the 
 gluon-fusion mechanism for pair-production of sleptons at the one-loop
 level. 
The corresponding production cross section, however, is at most of
 ${\cal O}(1)\,$fb for slepton and squark masses of
 ${\cal O}(100)\,$GeV.
We then extend our investigation to the gluon-fusion production of
 sleptons through the exchange of Higgs bosons.
The cross section is even smaller, of ${\cal O}(0.1)\,$fb, if the
 exchanged Higgs boson is considerably below the slepton-pair
 threshold, but it is enhanced when it is resonant.
It can reach ${\cal O}(10)\,$fb for the production of sleptons of
 same-chirality, exceeding these values for $\widetilde{\tau}$'s of
 opposite-chirality, even when chirality-mixing terms in the squark
 sector are vanishing.
The cross section can be further enhanced if these mixing terms are
 nonnegligible, providing a potentially interesting probe of the
 Higgs sector, in particular of parameters such as $A$, $\mu$, and
 $\tan\beta$.

\end{abstract}

\newpage 
\setcounter{page}{1}
\setlength{\parskip}{1.01ex}

\section{Introduction}
\label{sec:Intro}
%
It has been realized since quite some time that several extensions of 
 the Standard Model (SM) share similar signatures at the LHC, as, for
 example, that of jets, leptons, and missing transverse
 energy~\cite{SameSIGNALS}. 
This is because such extensions typically have 
 partners of the SM, with gauge quantum numbers equal to those
 of the SM particles~\cite{Review}.
Supersymmetric models, extra-dimensional ones, and little Higgs models 
 are indeed of this type. 
Among them, supersymmetric models stand out because the partners
 have spins that differ by $1/2$ from those of the SM particles, whereas
 in the other extensions have the same spins.
This has prompted several proposal of methods for determining the spin of
 heavy unstable particles~\cite{BIBLspinDET}.

Thus, supersymmetric models are also the only ones, at least so far, in
 which there exist irreducible renormalizable interactions of four matter
 partners, of type $\phi^{4}$.
Such interactions may, therefore, offer a new handle to discriminate
 Supersymmetry (SUSY) from other extensions of the SM.

The LHC being a hadronic machine, quartic couplings of sleptons, 
 $(\widetilde{l}^{\,\ast} \widetilde{l})^2$, 
 contribute only to processes whose cross sections are suppressed by
 many powers of the electroweak couplings.
The signals of quartic couplings of squarks,
 $(\widetilde{q}^{\,\ast} \widetilde{q})^2$, may not be easily
 disentangled from those due to a pair of squarks.
The first, simplest scrutiny of quartic couplings, therefore, may be that
 of couplings of two sleptons and two squarks
 $(\widetilde{q}^{\,\ast} \widetilde{q})
  (\widetilde{l}^{\,\ast} \widetilde{l})$.
Combined with the strong interaction vertices of squarks and gluons,
 these give rise, at the quantum level, to the effective couplings
 $gg\,\widetilde{l}^{\,\ast}\widetilde{l}$, and therefore to the
 gluon-initiated production of a pair of slepton.

Since Higgs bosons couple to both, a pair of fermions and a pair of
 scalars, reducible couplings of four SM partners, two quark partners
 and two lepton partners, can be built through the exchange of Higgs
 bosons also in models in which the partner spins do not differ from
 those of the SM particles.  
Thus, the gluon-fusion production of a pair of lepton partners is not 
 a special feature of supersymmetric models.
The possibility of discriminating SUSY through this type of studies
 strongly relies on the impact that the irreducible quartic couplings
 of supersymmetric models can have.

To the best of our knowledge, the one-loop process 
 $gg\,\widetilde{l}^{\,\ast}\widetilde{l}$ via quartic couplings, was
 never studied before. 
Surprisingly, very little exists in the literature also about
 the same process induced by the exchange of a Higgs boson,
 $gg\to H^0 \to \widetilde{l}^{\,\ast}\widetilde{l}$.
It was studied in Ref.~\cite{HIGGSmedGGone} for $H^0 =h$, the 
 lightest Higgs bosons. 
Since this has a mass clearly below the slepton-pair threshold, the 
 cross section was found to be very small.
The study of this process was put aside until the authors of 
 Ref.~\cite{HIGGSmedGGtwo} realized that the cross section could 
 increase if $H^0$ is one of the two heavier neutral Higgs bosons
 $H^0= H,A$. 
Neverheless, not much attention has been given to this proposal.

Although suppressed by a loop factor and by the coupling $\alpha_s$
 with respect to that of the standard Drell-Yan production
 $q \bar{q} \to \widetilde{l}^{\,\ast} \widetilde{l}$,
 the gluon-fusion cross sections may be favoured by the fact that, at
 the LHC, gluon-initiated processes may dominate over the
 quark-initiated ones. 
Given the importance of the quartic couplings that may be probed, it is
 worth studying them in detail.

The paper is organized as follows.
In Sec.~\ref{sec:QCXsec} we study the slepton pair production processes
 from gluon fusion induced by quartic couplings. 
In Sec.~\ref{sec:CompleteXsec} we give the complete one-loop level 
 results, including also the contribution from the Higgs
 boson-exchange.
We summarize our findings in Sec.~\ref{sec:Conclu}.

\section{Gluon-fusion production from quartic couplings}
\label{sec:QCXsec}
\subsection{Cross section}
\label{sec:QCXsecANAL}
%
The relevant quartic interactions in the scalar potential of the Minimal 
 Supersymmetric Standard Model (MSSM) consist of
 D-term interactions, proportional to gauge couplings squared, and
 F-terms interactions, proportional to Yukawa couplings squared.
These are respectively collected in:
\begin{equation}
 V_D \, \supset\, 
  d^{\,\widetilde{l}_i\widetilde{l}_i,\widetilde{q}_n\widetilde{q}_n}
 (\widetilde{q}_n^{\,\ast} \widetilde{q}_n)
 (\widetilde{l}_i^{\,\ast} \widetilde{l}_i),
\qquad\qquad
 V_F \, \supset\, 
  f^{\,\widetilde{l}_i\widetilde{l}_j,\widetilde{q}_m\widetilde{q}_n}
 (\widetilde{q}_m^{\,\ast} \widetilde{q}_n)
 (\widetilde{l}_i^{\,\ast} \widetilde{l}_j) +
 {\rm H.c.},  
\label{eq:VDVFcompact}
\end{equation}
 where $\widetilde{q}_n$ is any of the fields 
 $\widetilde{u}_L$, $\widetilde{d}_L$, 
 $\widetilde{u}_R$, $\widetilde{d}_R$, of any generation, and
 $\widetilde{l}_i$ is either the charged component of the SU(2)$_L$ doublet
 $(\widetilde{\nu}_L,\,\widetilde{l}_L)$ with hypercharge $-1/2$, or the
 SU(2)$_L$ singlet $\widetilde{l}_R$ with hypercharge $-1$, also of any
 generation.
The coefficients
 $d^{\,\widetilde{l}_i\widetilde{l}_i,\widetilde{q}_n\widetilde{q}_n}$
 and
 $f^{\,\widetilde{l}_i\widetilde{l}_j,\widetilde{q}_m\widetilde{q}_n}$
 are
 listed in Table~\ref{table:dANDfCOEFF}, where $g_1$ and $g_2$ are 
 the coupling constants of U(1)$_Y$ and SU(2)$_L$, with the
 normalization
 $g_1 \cos\theta_W =g_2 \sin \theta_W = e = \sqrt{4\pi \alpha}$.
The Yukawa coupling matrices, denoted by $h_U$, $h_D$, and $h_E$, are
 assumed to be diagonal and real, with only the $\{3,3\}$ element large
 enough to be relevant for this discussion.

These quartic couplings, together with the SU(3) interactions for
 squarks:
\begin{equation}
 K_{q_n^\ast q_n g (g)}   = 
  -g_S \, \widetilde{q}_n^{\,\ast} 
 \left(\!\!\frac{\lambda^a}{2}\!\right)
 \stackrel{\leftrightarrow}{\partial^{\,\mu}} 
          \widetilde{q}_n\,g^a_{\mu}
 \ + \ 
   g_S^2\,\widetilde{q}_n^{\,\ast} 
         \left(\!\!\frac{\lambda^a}{2}\frac{\lambda^b}{2}\!\right)
          \widetilde{q}_n \, g^a_{\mu} \,g^{b\,\mu},
\label{eq:KINsqglue}
\end{equation}
 allow a realization of the effective couplings
 $gg\,\widetilde{l}^{\,\ast}_i\widetilde{l}_j$ at the one-loop level,
 different from those due to the mediation of an s-channel Higgs boson.
In this expression, $g_S$ is the strong coupling and
 $(1/2)\{\lambda^a\}$ are the generators of SU(3).

\begin{table}[t]
\begin{center} 
\begin{tabular}{cccc} 
\hline\hline                     &&&\\[-1.9ex]       
$l_i^\ast l_j$                                       & 
$q_m^\ast q_n$                                       & 
$\qquad
  d^{\,\widetilde{l}_i\widetilde{l}_i,\widetilde{q}_n\widetilde{q}_n}$   & 
$\quad
  f^{\,\widetilde{l}_i\widetilde{l}_j,\widetilde{q}_m\widetilde{q}_n}
  \quad$                                           \\[1.01ex]
\hline                                               &&&\\[-1.9ex]     
$\quad \quad \widetilde{l}_L^\ast\widetilde{l}_L\quad \quad$  & 
$\widetilde{u}_L^\ast\widetilde{u}_L$                    & 
$\quad\displaystyle{-\frac{g_2^2}{4}} -\frac{g_1^2}{12}$ & 
$-$                                                 \\[1.4ex] 
$\quad \quad \widetilde{l}_L^\ast\widetilde{l}_L\quad \quad$  & 
$\widetilde{d}_L^\ast\widetilde{d}_L$                    & 
$\quad\displaystyle{+\frac{g_2^2}{4}} -\frac{g_1^2}{12}$ & 
$-$                                                 \\[1.4ex] 
$\widetilde{l}_L^\ast\widetilde{l}_L$                    & 
$\widetilde{d}_R^\ast\widetilde{d}_R$                    & 
$\quad -\displaystyle{\frac{g_1^2}{6}}$                  & 
$-$                                                 \\[1.4ex] 
$\widetilde{l}_L^\ast\widetilde{l}_L$                    & 
$\widetilde{u}_R^\ast\widetilde{u}_R$                    & 
$\quad+\displaystyle{\frac{g_1^2}{3}}$                   & 
$-$                                                 \\[1.4ex]
\hline                                               &&&\\[-1.9ex]     
$\widetilde{l}_R^\ast\widetilde{l}_R$                    & 
$\widetilde{u}_L^\ast\widetilde{u}_L$                    &  
$\quad+\displaystyle{\frac{g_1^2}{6}}$                   & 
$-$                                                 \\[1.4ex] 
$\widetilde{l}_R^\ast\widetilde{l}_R$                    & 
$\widetilde{d}_L^\ast\widetilde{d}_L$                    &  
$\quad+\displaystyle{\frac{g_1^2}{6}}$                   & 
$-$                                                 \\[1.4ex] 
$\widetilde{l}_R^\ast\widetilde{l}_R$                    & 
$\widetilde{d}_R^\ast\widetilde{d}_R$                    &  
$\quad+\displaystyle{\frac{g_1^2}{3}}$                   & 
$-$                                                 \\[1.4ex] 
$\widetilde{l}_R^\ast\widetilde{l}_R$                    & 
$\widetilde{u}_R^\ast\widetilde{u}_R$                    &  
$\quad-\displaystyle{\frac{2}{3}}g_1^2$                  & 
$-$                                                 \\[1.4ex] 
\hline                                               &&&\\[-1.9ex]     
$\widetilde{\tau}_L^\ast\widetilde{\tau}_R$              & 
$\widetilde{b}_R^\ast\widetilde{b}_L$                    &  
$\quad - \phantom{\frac{2}{3}}$                          & 
$ (h_{D})_{3,3}(h_{E})_{3,3}$                       \\[1.0001ex]       
\hline\hline
\end{tabular} 
\caption{\small{\it Coefficients
 $d^{\,\widetilde{l}_i\widetilde{l}_i,\widetilde{q}_n\widetilde{q}_n}$
 and
 $f^{\,\widetilde{l}_i\widetilde{l}_j,\widetilde{q}_m\widetilde{q}_n}$
 for quartic squark--charged slepton interactions, from D- and F-terms.
 The conjugated term $(\widetilde{\tau}_R^\ast\widetilde{\tau}_L)
  (\widetilde{b}_L^\ast\widetilde{b}_R)$ also exists, but given our
 approximation for the Yukawa couplings, it has an F-term coefficient
 equal to that of $(\widetilde{\tau}_L^\ast\widetilde{\tau}_R)
  (\widetilde{b}_R^\ast\widetilde{b}_L)$.
}}
\label{table:dANDfCOEFF}
\end{center}
\end{table}

\begin{figure}[t] 
\begin{center} 
\epsfxsize= 5.3 cm 
\leavevmode 
\epsfbox{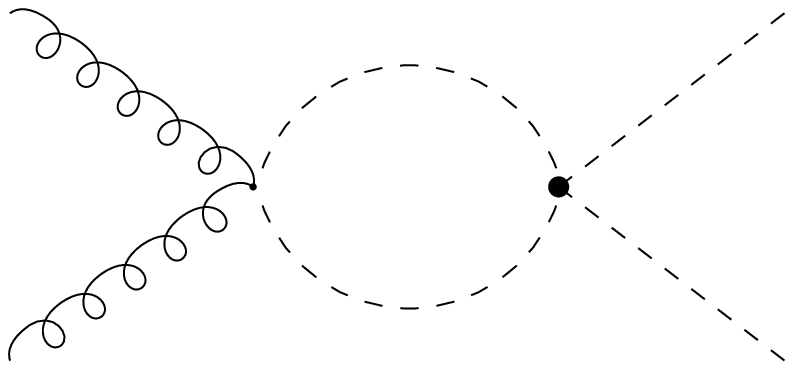} 
\put(-82,37){$\widetilde{q}_n$}
\put(-13,44){$\widetilde{l}^\ast_i$}
\put(-13,5){$\widetilde{l}_j$}
\hspace*{2truecm} 
\epsfxsize= 5.3 cm 
\epsfbox{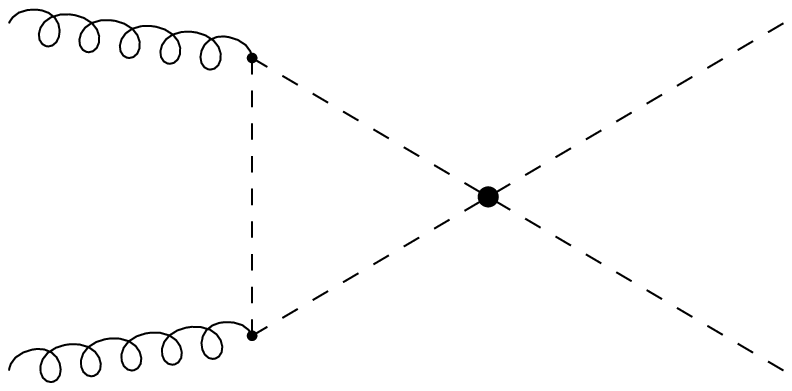} 
\put(-100,26){$\widetilde{q}_n$}
\put(-13,44){$\widetilde{l}^\ast_i$}
\put(-13,5){$\widetilde{l}_j$}
\end{center} 
\caption{\small{\it Diagrams contributing to the effective couplings
 $gg\,\widetilde{l}_i^\ast\widetilde{l}_j$.
 The big dot indicates a squark--slepton quartic coupling.
}} 
\label{figDIAG:QCloop} 
\end{figure}

Examples of loop diagrams giving rise to the effective interactions
 $gg\,\widetilde{l}_i^\ast\widetilde{l}_j$ through quartic couplings are
 given in Fig.~\ref{figDIAG:QCloop}. 
The corresponding parton-level cross section is
\begin{equation}
 \widehat{\sigma}_{\,\rm QC}
 (gg\to \widetilde{l}_i^\ast \widetilde{l}_j) =  
 \frac{\alpha_s^2}{(16\pi)^3} \, 
 \frac{\lambda^{1/2}(\hat{s},\widetilde{m}_{l_i},\widetilde{m}_{l_j})}
      {\hat{s}} \,
 \left\vert \widetilde{A}_{\,\rm QC}^{\,\widetilde{l}_i \widetilde{l}_j}
 \right\vert^2,
\label{eq:D-QCXsec}
\end{equation}
 where 
 $\lambda^{1/2}(\hat{s},m_1,m_2) =\!
   \left[1\! -\! {(m_1\!\!+\!\!m_2)^2}/{\hat{s}} \right]^{1/2}\!
   \left[1\! -\! {(m_1\!\!-\!\!m_2)^2}/{\hat{s}} \right]^{1/2}$, 
 for the production of two particles with mass $m_1$ and $m_2$, 
 or simply
 $\lambda^{1/2}(\hat{s},m) =\!\left[1\!-\! 4 m^2/\hat{s}\right]^{1/2}$,
 the usual $\beta$ factor, if $m_1=m_2=m$.
A dependence on the produced slepton masses is also hidden in 
 $\hat{s}$, as the minimum value that $\hat{s}$ can acquire is 
 $\hat{s}_{\rm min}=(\widetilde{m}_{l_i}+\widetilde{m}_{l_j})^2$. 
The truncated amplitude
 $\widetilde{A}_{\,\rm QC}^{\,\widetilde{l}_i \widetilde{l}_j}$ is  calculated
 hereafter.

Since the strong interactions do not change chirality, contributions to
 $\widetilde{A}_{\,\rm QC}^{\,\widetilde{l}_i \widetilde{l}_j}$ from
 F-terms are possible only for nonvanishing chirality-mixing terms in
 the $\widetilde{b}$ squark sector.
These are included in the mass-mixing terms of the scalar potential:
\begin{equation}
 V_{\rm ch-mix} \, \supset \, 
 \widetilde{u}_R^\ast \widetilde{m}_{U,{\rm RL}}^2 \widetilde{u}_L+ 
 \widetilde{d}_R^\ast \widetilde{m}_{D,{\rm RL}}^2 \widetilde{d}_L+ 
 \widetilde{l}_R^\ast \widetilde{m}_{E,{\rm RL}}^2 \widetilde{l}_L+ 
 {\rm H.c.},
\label{eq:trilinear}
\end{equation}
 which contribute, together with those in the chirality-conserving
 soft mass terms:
\begin{equation}
 V^{\rm soft}_{\rm 2} = 
 (\widetilde{u}_L^\ast,\,\widetilde{d}_L^\ast)
 \, \widetilde{m}_{Q}^2
 \left(\!\!\! \begin{array}{c} \widetilde{u}_L \\ \widetilde{d}_L 
              \end{array}\!\!\!\right) + 
 \widetilde{u}_R^\ast \widetilde{m}_{U}^2 \,\widetilde{u}_R + 
 \widetilde{d}_R^\ast \widetilde{m}_{D}^2 \,\widetilde{d}_R + 
 (\widetilde{\nu}_L^\ast,\,\widetilde{l}_L^\ast)
 \, \widetilde{m}_{L}^2 
 \left(\!\!\! \begin{array}{c} \widetilde{\nu}_L \\ \widetilde{l}_L 
              \end{array}\!\!\!\right) + 
 \widetilde{l}_R^\ast \widetilde{m}_{E}^2 \,\widetilde{l}_R , 
\label{eq:softmass}
\end{equation}
 to the up- and down-squark as well as slepton mass matrices. 
In our discussion, the parameters
 $\widetilde{m}_{Q}^2$, $\widetilde{m}_{U}^2$,
 $\widetilde{m}_{D}^2$, $\widetilde{m}_{L}^2$,
 $\widetilde{m}_{E}^2$ are $3\times 3$ diagonal Hermitian matrices
 and the chirality-mixing parameters
 $\widetilde{m}_{U,{\rm RL}}^2$, $\widetilde{m}_{D,{\rm RL}}^2$, 
 $\widetilde{m}_{E,{\rm RL}}^2$ are
\begin{equation}
 \widetilde{m}_{U,{\rm RL}}^2 = 
 \frac{\left(\!v_u A_{U\!}-\!\mu v_d\!\right) h_U}
      {\sqrt{2}}, 
\quad
 \widetilde{m}_{D,{\rm RL}}^2 = 
 \frac{\left(\!v_d A_{D\!}-\!\mu v_u\!\right) h_D}
      {\sqrt{2}}, 
\quad
 \widetilde{m}_{E,{\rm RL}}^2 = 
 \frac{\left(\!v_d A_{E\!}-\!\mu v_u\!\right) h_E}
      {\sqrt{2}}, 
\label{eq:LRmixingMass}
\end{equation}
 where $v_d$, $v_u$ are the vacuum expectation values of the neutral
 component in the two Higgs doublets $H_d$ and $H_u$, with
 $\tan \beta = v_u/v_d$ and $v^2=v_d^2+v_u^2$. 
(Our normalization is such that $M_W = g v /2$.)
The parameters ${A}_U$, ${A}_D$, and ${A}_E$ are the massive couplings
 of the trilinear soft terms holomorphic in the Higgs fields. 
We neglect also here any possible flavor mixing, {\it i.e.}
 ${A}_U$, ${A}_D$, and ${A}_E$ are, for our purposes, diagonal
 matrices, with real elements, and the terms in $V_{\rm ch-mix}$ are
 assumed to be sizable only for third-generation squarks.

If all chirality-mixing terms in $V_{\rm ch-mix}$ are negligible
 with respect to the soft mass squared in $V^{\rm soft}_{\rm 2}$,
 in which case current eigenstates coincide to a good approximation
 with mass eigenstates, only same-chirality sleptons can be produced, 
 induced by D-terms only. 
The corresponding truncated amplitude 
 $\widetilde{A}_{\,\rm QC}^{\,l_i l_i}$ has the simple form:
\begin{equation} 
 \widetilde{A}_{\,\rm QC}^{\,\widetilde{l}_i \widetilde{l}_i}  = 
  - \sum_{q=u,d}               
 \left[
   d^{\,\widetilde{l}_i\widetilde{l}_i,\widetilde{q}_L \widetilde{q}_L}
   S(\hat{s};\widetilde{m}_{q_L}^2) + 
   d^{\,\widetilde{l}_i\widetilde{l}_i,\widetilde{q}_R \widetilde{q}_R}
   S(\hat{s};\widetilde{m}_{q_R}^2)\right],  
\label{eq:D-truncAMPL}
\end{equation}
 where $l_i$ labels the type of produced sleptons, $l_i =\{l_L,l_R\}$, 
 with $l=e,\mu,\tau$, and the sum in $q$ is understood to be over all
 three generations: $u\to u,c,t$, and $d\to d,s,b$.
The loop function $S(\hat{s};\widetilde{m}_{q_n}^2)$ depends on $\hat{s}$
 and $\widetilde{m}_{q_n}^2$ only through their ratio:
\begin{equation}
 S(\hat{s};m^2) =   -1 + \tau f(\tau),   \hspace*{1truecm}  
 \tau \equiv 4 m^2/\hat{s},
\label{eq:functionS} 
\end{equation} 
 where $f(\tau)$ is defined as:
\begin{equation} 
 f(\tau) =
 \left\{ \begin{array}{ll} 
 \displaystyle    
  - \frac{1}{4} \left[\log  
 \left(\frac{1+\sqrt{1\!-\!\tau}}{1-\sqrt{1\!-\!\tau}}\right) 
  - i \pi   \right]^2   
         &   \quad \tau  < 1   \\[1.7ex]          
 \displaystyle + \arcsin^2\left(\frac{1}{\sqrt{\tau}}\right) 
         &   \quad \tau \ge 1  
\end{array} \right. .
\label{eq:functionf} 
\end{equation}  
As mentioned already, there is an implicit dependence on the mass of the
 produced sleptons, due to the fact that $\hat{s}_{\rm min}$, the minimal
 value that $\hat{s}$ can have, depends on it.

A quick inspection of Table~\ref{table:dANDfCOEFF} shows that:
\begin{equation}
 \sum_{\widetilde{q}_n}
 d^{\,\widetilde{l}_L\widetilde{l}_L,\widetilde{q}_n\widetilde{q}_n}
 \ =\ 
 \sum_{\widetilde{q}_n}
 d^{\,\widetilde{l}_R\widetilde{l}_R,\widetilde{q}_n\widetilde{q}_n}
 \ =\ 0,
\label{eq:chSUMRULE1}
\end{equation}
 for each generation of squarks.
This is no surprise, as, for each generation, the sums of the isospin
 and the weak hypercharge of all quarks and leptons vanish in the SM. 
The consequence of these sum rules is that, in the limit of degenerate
 squark masses,
 $\widetilde{A}_{\,\rm QC}^{\,\widetilde{l}_L \widetilde{l}_L}$ and 
 $\widetilde{A}_{\,\rm QC}^{\,\widetilde{l}_R \widetilde{l}_R}$ vanish
 identically.

A complete degeneracy of all squarks of the same generation, is however
 unlikely. 
Sizable differences between the soft mass of the right-handed and 
 left-handed squarks at low energy are present in most supersymmetric
 models.
To give an example, in models with universal boundary conditions for 
 soft masses at some high scale, typically the
 Planck mass or the scale of grand unification, a sufficient splitting
 between the soft mass of the right- and left-handed squarks at the 
 electroweak scale is induced by the fact
 that, contrary to $\widetilde{m}_{U}^2$ and
 $\widetilde{m}_{D}^2$, $\widetilde{m}_{Q}^2$ feels the 
 effect of the SU(2)$_L$ gaugino during the downward evolution.
In typical gauge-mediation models, with a weekly coupled messenger 
 sector, a splitting is already present at the scale
 of SUSY breaking and may be further enlarged by a subsequent evolution
 down to the electroweak scale. 
Thus, the value of
 $\widetilde{A}_{\,\rm QC}^{\,\widetilde{l}_i \widetilde{l}_i}$ can be easily
 lifted from zero.

If some of the terms in $V_{\rm ch-mix}$ cannot be neglected, the
 expression for $\widetilde{A}_{\,\rm QC}^{\,\widetilde{l}_i \widetilde{l}_i}$ in 
 Eq.~(\ref{eq:D-truncAMPL}) is not valid. 
Indeed, even for small values of $\tan \beta$, the above approximation
 may fail for $\widetilde{t}$ squark states, but also for other squarks
 states, for large values of $A_U$ or $A_D$.
In this case, the cross section must be expressed in terms of the
 sfermion mass eigenstates:
%
 $\widetilde{f}_1 = \widetilde{f}_L \cos \theta_{f} + 
                    \widetilde{f}_R \sin \theta_{f}$ ,  
 $\widetilde{f}_2 = \widetilde{f}_L \sin \theta_{f} -  
                    \widetilde{f}_R \cos \theta_{f}$ , 
 and the mixing angles $\theta_f$, whose sine and cosine are often
 abbreviated in the following as $s_f$ and $c_f$, respectively.
That for $\widetilde{t}$ squarks is defined by the relation:
\begin{equation}
 \cos 2 \theta_t  = 
  \frac{(\widetilde{m}_{\bar{U}}^2)_{3,3}-(\widetilde{m}_{Q}^2)_{3,3}
          +(\Delta_D)_u} 
        {\widetilde{m}_{t_2}^2 - \widetilde{m}_{t_1}^2},
\label{eq:SQUPcos2theta}
\end{equation} 
 where the numerator is the difference between the often called
 right-right and left-left diagonal entries in the $\widetilde{t}$ squark
 mass matrix.
In particular, $(\Delta_D)_u$ is the difference of the D-term
 contributions to these entries.
A similar formula holds for $\widetilde{b}$ squarks, with the
 replacements
 $(\widetilde{m}_{\bar{U}}^2)_{3,3}\to (\widetilde{m}_{\bar{D}}^2)_{3,3}$,
 $\widetilde{m}_{t_i}^2\to \widetilde{m}_{b_i}^2$ ($i=1,2$), and 
 $(\Delta_D)_u \to (\Delta_D)_d$. 
It can be easily adapted also to the case of $\widetilde{\tau}$
 sleptons. 
The explicit expressions for $(\Delta_D)_u$, $(\Delta_D)_d$, and
 $(\Delta_D)_l$ can be found in any pedagogical review of SUSY.
 (See for example Ref.~\cite{HIKASA}.)
For $\tan \beta \gsim 3$, which is always assumed in this paper, these
 quantities can be approximated as
\begin{equation}
 (\Delta_D)_u =  +\frac{1}{6} M_Z^2,
\qquad \
 (\Delta_D)_d =  -\frac{1}{3} M_Z^2, 
\qquad \
 (\Delta_D)_l =  0.
\label{eq:DtermDIFF}
\end{equation}
Notice that the usual two-fold ambiguity in the definition of 
 $\cos 2\theta_t$ has been implicitly removed with the assumption 
 $(\widetilde{m}_{\bar{U}}^2)_{3,3} <
  (\widetilde{m}_{Q}^2)_{3,3}-(1/6)M_Z^2$.
In the limit of vanishing chirality-mixing terms for a sfermion 
 $f$, $\cos\theta_{f}\to 0$,
 $\widetilde{f}_1 \to \widetilde{f}_R$, and 
 $\widetilde{f}_2 \to \widetilde{f}_L$.

The generalization of
 $\widetilde{A}_{\,\rm QC}^{\,\widetilde{l}_i \widetilde{l}_i}$ to the
 case of nonnegligible chirality-mixing terms in the squark sector is
 then straightforward:
\begin{eqnarray}
\widetilde{A}_{\,\rm QC}^{\,\widetilde{l}_i \widetilde{l}_i} & = & 
 -\sum_{q=u,d}
  \left[
 \left(
 d^{\,\widetilde{l}_i \widetilde{l}_i,\widetilde{q}_L \widetilde{q}_L}
 c_q^2 +
 d^{\,\widetilde{l}_i \widetilde{l}_i,\widetilde{q}_R \widetilde{q}_R}
 s_q^2
 \right)   S(\hat{s};m_{\widetilde{q}_1}^2) 
  \right. + 
\nonumber\\[-0.9ex]                  &   & 
  \left. \hspace*{1.5truecm}
 \left(
 d^{\,\widetilde{l}_i \widetilde{l}_i,\widetilde{q}_L \widetilde{q}_L}
 s_q^2 +
 d^{\,\widetilde{l}_i \widetilde{l}_i,\widetilde{q}_R \widetilde{q}_R}
 c_q^2
 \right)   S(\hat{s};m_{\widetilde{q}_2}^2) 
  \right].
\label{eq:genQCXsec}
\end{eqnarray}
These truncated amplitudes apply to the production of pairs of
 same-chirality sleptons of first and second generations, for any value
 of $\tan \beta$, and to the production of a pair of same-chirality
 $\widetilde{\tau}$'s, for small values of $\tan \beta$.

When $\tan \beta$ is large, however, the chirality-mixing terms in the
 $\widetilde{\tau}$ sector may not be small. 
(See also Sec.~\ref{sec:QCXsecNUMER}.)
Thus, $\widetilde{\tau}_L$ and $\widetilde{\tau}_R$ are not mass 
 eigenstates anymore.
The truncated amplitudes for the production of the three final states
 $\widetilde{\tau}_1\widetilde{\tau}_1$, 
 $\widetilde{\tau}_2\widetilde{\tau}_2$,
 and $\widetilde{\tau}_1\widetilde{\tau}_2$ are
\begin{eqnarray}
 \widetilde{A}_{\,\rm QC}^{\,\widetilde{\tau}_1 \widetilde{\tau}_1} 
& = &  
 c_\tau^2 
 \widetilde{A}_{\,\rm QC}^{\,\widetilde{\tau}_L \widetilde{\tau}_L} +
 s_\tau^2 
 \widetilde{A}_{\,\rm QC}^{\,\widetilde{\tau}_R \widetilde{\tau}_R},
\nonumber\\[1.1ex]
 \widetilde{A}_{\,\rm QC}^{\,\widetilde{\tau}_2 \widetilde{\tau}_2} 
& = &  
 s_\tau^2 
 \widetilde{A}_{\,\rm QC}^{\,\widetilde{\tau}_L \widetilde{\tau}_L} +
 c_\tau^2 
 \widetilde{A}_{\,\rm QC}^{\,\widetilde{\tau}_R \widetilde{\tau}_R},
\nonumber\\[1.001ex]
 \widetilde{A}_{\,\rm QC}^{\,\widetilde{\tau}_1 \widetilde{\tau}_2}
& = &  
( s_\tau c_\tau) \left[
 \widetilde{A}_{\,\rm QC}^{\,\widetilde{\tau}_L \widetilde{\tau}_L} -
 \widetilde{A}_{\,\rm QC}^{\,\widetilde{\tau}_R \widetilde{\tau}_R}
           \right]. 
\label{eq:genQCXsecGENslep} 
\end{eqnarray}

If among the nonvanishing terms of $V_{\rm ch-mix}$ there is also 
 $(\widetilde{m}^2_{D,{\rm RL}})_{3,3}$, F-terms induce a contribution
 to the cross section for the production of opposite-chirality sleptons. 
This is of relevance only for large values of $\tan \beta$ and for
 the production of a pair of $\widetilde{\tau}$ sleptons.
Notice that no chirality-mixing terms in the $\widetilde{\tau}$ 
 sector is necessary for this contribution to exist. 
As mentioned earlier, however, since $\tan \beta$ is large, it is hard to
 imagine that such mixing terms really vanish.
The truncated amplitudes for the production of the three final states
 $\widetilde{\tau}_1\widetilde{\tau}_1$, 
 $\widetilde{\tau}_2\widetilde{\tau}_2$,
 and $\widetilde{\tau}_1\widetilde{\tau}_2$ are now
\begin{eqnarray}
 \widetilde{A}_{\,\rm QC}^{\,\widetilde{\tau}_1 \widetilde{\tau}_1}       
& = &
+ 2 (s_\tau c_\tau)
\widetilde{A}_{\,\rm QC}^{\,\widetilde{\tau}_L \widetilde{\tau}_R}, 
\nonumber\\
 \widetilde{A}_{\,\rm QC}^{\,\widetilde{\tau}_2 \widetilde{\tau}_2}       
& = &
- 2 (s_\tau c_\tau)
\widetilde{A}_{\,\rm QC}^{\,\widetilde{\tau}_L \widetilde{\tau}_R}, 
\nonumber\\
 \widetilde{A}_{\,\rm QC}^{\,\widetilde{\tau}_1 \widetilde{\tau}_2}       
& = &
- (c^2_\tau - s^2_\tau )
 \widetilde{A}_{\,\rm QC}^{\,\widetilde{\tau}_L \widetilde{\tau}_R}. 
\label{eq:F-truncAMPL}
\end{eqnarray}
Here,
 $\widetilde{A}_{\,\rm QC}^{\,\widetilde{\tau}_L \widetilde{\tau}_R}$ 
 is the truncated amplitude obtained from the F-term quartic couplings
 in the limit of vanishing chirality-mixing terms in the
 $\widetilde{\tau}$ sector:
\begin{equation}
 \widetilde{A}_{\,\rm QC}^{\,\widetilde{\tau}_L \widetilde{\tau}_R}
 = 
 -\frac{g_2^2}{2} \frac{m_b m_\tau}{M_W^2} 
  \frac{1}{\cos^2\beta} \, (s_b c_b) 
  \left[ S(\hat{s};\widetilde{m}_{b_1}^2) -
         S(\hat{s};\widetilde{m}_{b_2}^2) \right] .
\label{eq:F-LRtruncAMPL}
\end{equation}
Notice that these truncated amplitudes scale as $\tan^2 \beta$, for any
 value of $\tan \beta\gsim 3$, and are also very sensitive to
 $\vert (\widetilde{m}^2_{D,{\rm RL}})_{3,3}\vert$.
Since
 $\vert \cos \theta_b \sin \theta_b \vert \sim 
  \vert (\widetilde{m}^2_{D,{\rm RL}})_{3,3}\vert/
 (\widetilde{m}_{b_2}^2-\widetilde{m}_{b_1}^2)$, they roughly scale linearly
 in this parameter.

At large $\tan \beta$, when both contributions from D- and F-term
 couplings exist, the truncated amplitudes 
 $ \widetilde{A}_{\,\rm QC}^{\, \widetilde{\tau}_1 \widetilde{\tau}_1}$, 
 $ \widetilde{A}_{\,\rm QC}^{\,\widetilde{\tau}_2 \widetilde{\tau}_2}$,
 and
 $ \widetilde{A}_{\,\rm QC}^{\,\widetilde{\tau}_1 \widetilde{\tau}_2}$
 are obtained by summing up those in Eqs.~(\ref{eq:genQCXsecGENslep})
 and~(\ref{eq:F-truncAMPL}).

\subsection{Results}
\label{sec:QCXsecNUMER}
%
We are now in a position to show some results, focussing on spectra that
 showcase separately the contribution from F- and D-terms.

\subsubsection{D-term contributions}
\label{sec:QCXsecNumD}
%
For the first spectrum, or spectrum~{\bf A}, we choose all left-handed
 squarks and the right-handed ones of down-type to be heavy,
 {\it i.e.} $1\,$TeV, whereas the right-handed squarks of up-type and
 all sleptons are light.
The values chosen here for the massive parameters in the down-squark
 sector are somewhat extreme, given how light the other squarks are.
Indeed, it does not need to be so large, as the contribution from
 loops with squark heavier than a few hundred GeV decouples quite
 rapidly.
The chirality-mixing terms $(\widetilde{m}_{U,{\rm RL}}^2)_{ii}$ and 
 $(\widetilde{m}_{E,{\rm RL}}^2)_{ii}$, with $i=1,2,3$, are assumed to be
 vanishing, whereas  the terms $(\widetilde{m}_{D,{\rm RL}}^2)_{ii}$ do not
 need to be specified, as the down-squarks are decoupled from our problem.
Thus, the diagonal entries in the up-squark mass matrix and in the
 slepton mass matrices are mass eigenvalues.
We denote the up-squarks relevant for our calculation  
 $\widetilde{m}_{u_R}$, $\widetilde{m}_{c_R}$, $\widetilde{m}_{t_R}$.
Under the assumption that all three entries in the diagonal matrix
 $\widetilde{m}_{\bar{U}}^2$ are equal,
 $\widetilde{m}_{u_R}$, $\widetilde{m}_{c_R}$ are given by 
 $\widetilde{m}_{q}^2 \simeq (\widetilde{m}_{\bar{U}}^2)_{ii} -M_Z^2/6$, for 
 $i=1,2$ 
 (see Eq.~(\ref{eq:DtermDIFF})), whereas $\widetilde{m}_{t_R}$ receives
 also the F-term contribution $m_t^2$.

In summary, this spectrum is specified as follows:
\begin{equation}
\begin{array}{ll}
\hspace*{-0.5truecm}
\underline{{\rm spectrum} \ \mbox{\boldmath{${\rm A}$}}:} 
&  
\\[1.1ex]
{\rm squarks}: 
& 
\left\{
\begin{array}{ll}
 (\widetilde{m}_{Q}^2)_{ii} \ = (\widetilde{m}_{\bar{D}}^2)_{ii}
                         \ = (1\, {\rm TeV})^2
& \ \quad (i=1,2,3),
\\[1.001ex]
 (\widetilde{m}_{U,{\rm RL}}^2)_{ii} \simeq 0 
& \ \quad (i=1,2,3),
\\[1.01ex]
  \widetilde{m}_{u_R}^2   \ = \, \widetilde{m}_{c_R}^2  \ \equiv \
  \widetilde{m}_{q}^2 \ \ll  (\widetilde{m}_{Q}^2)_{ii},
& 
\\
  \widetilde{m}_{t_R}^2\, \ = (\widetilde{m}_{q})^2 +\ m_t^2 ,
& 
\end{array}
\right.
\\[8ex]
{\rm sleptons}: 
& 
\, 
\left\{
\begin{array}{ll}
 (\widetilde{m}_{E,{\rm RL}}^2)_{ii}  \simeq   0 
& \quad \qquad \qquad\  (i=1,2,3),
\\
\,\widetilde{m}_{e_L} \ = \,\widetilde{m}_{\mu_L}  \  =  \
  \widetilde{m}_{\tau_L},
&  
\\
\,\widetilde{m}_{e_R} \ = \,\widetilde{m}_{\mu_R}  \  =  \
  \widetilde{m}_{\tau_R}.
&  
\end{array}
\right.
\label{eq:spectrumA}
\end{array}
\end{equation}

With this spectrum, only the production of same-chirality sleptons is
 possible, induced by the D-term contributions to the quartic couplings,
 $d^{\,\widetilde{l}_i \widetilde{l}_i,\widetilde{u}_R \widetilde{u}_R}$.
We allow values for $\widetilde{m}_{q}$ in the interval $[100,200]\,$GeV.
Notice that the mass of the third generation right-handed
 $\widetilde{t}$ squarks ranges approximately between $200$ and
 $260\,$GeV.
Since it is heavier, it gives a contribution to the cross sections 
 smaller than that of the other right-handed up squarks.

\begin{figure}[t] 
\begin{center} 
\epsfxsize= 12.5cm 
\leavevmode 
\epsfbox{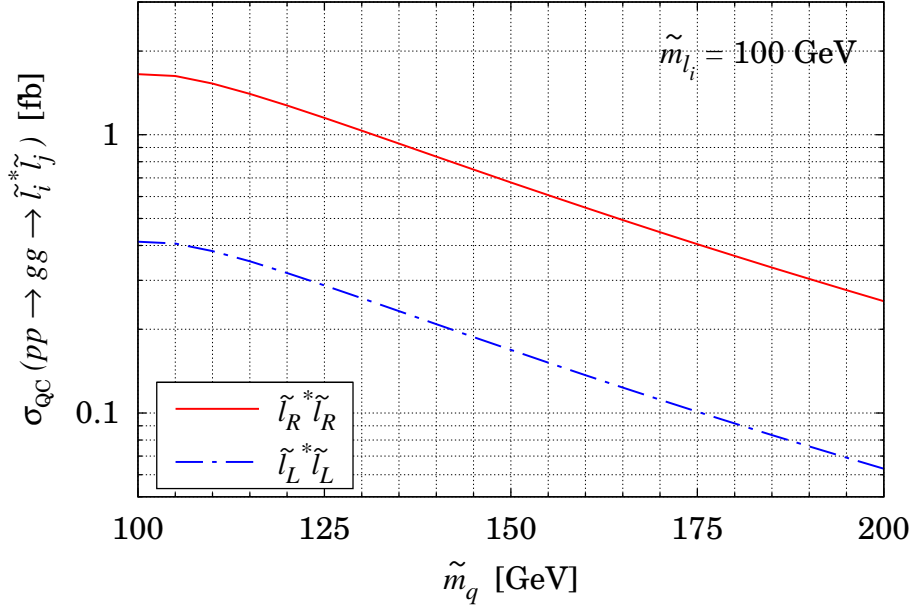} 
\vspace*{-0.5truecm}
\end{center} 
\caption{\small{\it Quartic-coupling induced cross sections for the
  production of a pair of $100\,$GeV, same-chirality sleptons obtained
  for the spectrum~{\bf A}~(see Eq.~(\ref{eq:spectrumA})).
 The upper (red solid) and lower (blue dot-dashed) line correspond to
  right-handed and left-handed sleptons, respectively.
 These cross sections apply to the production of first and second
  generation sleptons for any value of $\tan \beta$, of third
  generation, only for low $\tan beta$~(see discussion in the text).  
}} 
\label{figPLOT:QCxscA} 
\end{figure}

The production cross sections are shown in Fig.~\ref{figPLOT:QCxscA} as
 a function of $\widetilde{m}_{q}$.
The upper line (the red solid line) corresponds to the production of 
 right-handed sleptons, which decreases from 1.6$\,$fb to 0.25$\,$fb
 in the mass range considered. 
The lower line (the blue dot-dashed line) corresponds to the production
 of left-handed sleptons, which is smaller by a factor four.
This is due to the fact that the hypercharge of
 right-handed leptons is twice as large as that of left-handed ones,
 $d^{\,\widetilde{l}_R \widetilde{l}_R,\widetilde{u}_R \widetilde{u}_R}/
  d^{\,\widetilde{l}_L \widetilde{l}_L,\widetilde{u}_R \widetilde{u}_R}
 =-2$.

These, as well as all other cross sections shown in this paper are
 obtained using the CTEQ6L parton distribution functions~\cite{CTEQ},
 with factorization and renormalization scales fixed at $\sqrt{\hat{s}}$
 and the adaptive Monte Carlo integration program
 {\sf BASES}~\cite{BASES}.

The cross sections shown in this figure apply to the production of
 first and second generation sleptons, for any value of $\tan \beta$,
 and to the production of a $\widetilde{\tau}$'s pair only for low
 $\tan \beta$.
This is due to the fact that for large values of $\tan \beta$ the
 Yukawa coupling of the $\tau$ lepton is nonnegligible, and the 
 condition $(\widetilde{m}_{E,{\rm RL}}^2)_{33}\simeq 0$ can only be realized
 by tuning $(A_E)_{3,3}$ to cancel $\mu\tan \beta$.
Large values of $\vert(A_E)_{3,3}\vert$, however, may have to be required
 for this cancellation, possibly producing charged minima lower than the
 electroweak one~\cite{CHH}. 
Of course it is still possible to assume that the local
 charge-preserving vacuum is metastable, with a lifetime greater than
 the age of the Universe~\cite{LargeAtunn}.
Nevertheless, given how light the slepton spectrum is, also the
 metastability of the charge-preserving vacuum may be jeopardized by a
 large value of $\tan \beta$.
Therefore, it is safe to say that the cross section in this figure
 apply to the production of a pair of $\widetilde{\tau}$'s only in the
 case of small values of $\tan \beta$.

It should be noted here that the spectrum~{\bf A} has been tailored to
 maximize the value of these two production cross sections.
On one side, the lightness of some of the squarks in this spectrum is 
 extreme. 
On the other side, the choice of the up-type right-handed squarks as the 
 light ones, has also contributed to increase the cross sections.
The choice of the right-handed down squarks as light would have given 
 smaller cross section,  due to the fact that the corresponding quartic
 couplings have a ratio
 $ d^{\,\widetilde{l}_i \widetilde{l}_i,\widetilde{u}_R \widetilde{u}_R}/
   d^{\,\widetilde{l}_i \widetilde{l}_i,\widetilde{d}_R \widetilde{d}_R} =-2$.
A similar suppression factor is found in the situation in which the
 left-handed squarks are light, whereas the right-handed ones decouple
 from the problem. 
Due to SU(2)$_L$ gauge invariance, the equality of the eigenvalues 
 $\widetilde{m}_{u_L}^2 = \widetilde{m}_{d_L}^2$ holds to a very good
 approximation, except for the third generation.
Thus, the two squarks $\widetilde{u}_L$ and $\widetilde{d}_L$ exchanged
 in the loop contribute only with the U(1)$_Y$ part of their couplings 
 $d^{\,\widetilde{l}_i \widetilde{l}_i,\widetilde{u}_L \widetilde{u}_L}$
 and
 $d^{\,\widetilde{l}_i \widetilde{l}_i,\widetilde{d}_L \widetilde{d}_L}$.
As Table~\ref{table:dANDfCOEFF} shows, they contribute to the cross
 sections as much as one right-handed down squark of equal mass does.

The cross sections shown in this figure, therefore, can be considered as
 an upper bound for the value of quartic-coupling induced cross
 sections. 
One could try to increase this bound  by taking $(\widetilde{m}_{Q}^2)_{33}$
 smaller than $(\widetilde{m}_{Q}^2)_{ii}$ ($i=1,2$), and therefore
 lowering the value of $\widetilde{m}_{t_R}$.
The gain is, however, quite small, as $\widetilde{m}_{t_R}$ cannot get
 down to the values of $\widetilde{m}_{u_R}$ and $\widetilde{m}_{c_R}$.
A substantial chirality-mixing term in the $\widetilde{t}$ squark
 sector, could lower the value of $\widetilde{m}_{t_1}$.
As Eq.~(\ref{eq:genQCXsecGENslep}) shows, the contribution of such a 
 lighter $\widetilde{t}_1$ to the cross section would, however, get
 penalized by the mixing angle $\sin \theta_t$.

\subsubsection{F-term contributions}
\label{sec:QCXsecNumF}
%
Large chirality-mixing terms in the $\widetilde{b}$-squark sector are 
 needed to have a nonvanishing contribution to the production of
 opposite-chirality $\widetilde{\tau}$ sleptons from the F-term 
 $(h_D)_{33} (h_E)_{33}$.
Indeed, the corresponding cross section scales slightly more strongly
 than $r_b^2$, where $r_b$ is defined as 
 $\vert (\widetilde{m}^2_{D,{\rm RL}})_{3,3}\vert =
  r_b \widetilde{m}_{b_1}^2$, and 
 as $(\sin 2\theta_b)^2$ (see Eq.~(\ref{eq:F-LRtruncAMPL})).
Thus, both these parameters need to be maximized, while keeping 
 the $\widetilde{b}$-squark and the $\widetilde{\tau}$-slepton
 relatively light, in order to explore an upper bound for this cross
 section. 
Moreover, given its scaling like $(\tan \beta)^4$, also $\tan \beta$
 needs to be large.
As a consequence, the $\widetilde{\tau}$-slepton sector needs to be 
 strongly mixed as well.

We start by requiring 
 $\cos\theta_b =\sin\theta_b =1/\sqrt{2}$, so that $\cos 2 \theta_b=1$ 
 (see Eq.~(\ref{eq:SQUPcos2theta})).
This requirement fixes also
 $\widetilde{m}_{b_2}^2$ in terms of $\widetilde{m}_{b_1}^2$ and $r_b$. 
It is, indeed, $\widetilde{m}_{b_2}^2=(1+2r_b)\widetilde{m}_{b_1}^2$.
Strictly speaking, only the $\widetilde{b}$-squark sector needs to be
 specified for this cross section.
Nevertheless, if we require also a maximal mixing in the $\widetilde{t}$ 
 sector, {\i.e.} $\cos\theta_t =\sin\theta_t =1/\sqrt{2}$, because of
 SU(2)$_L$ gauge invariance, the above numbers specify also the diagonal
 entries in the $\widetilde{t}$ squark mass matrix.
As a consequence, $\widetilde{m}_{t_1}^2$ and $\widetilde{m}_{t_2}^2$ are 
 fixed in terms of $\widetilde{m}_{b_1}^2$, $r_b$ and $r_t$,  
 a parameter analogous to $r_b$:
 $\vert (\widetilde{m}^2_{U,{\rm RL}})_{3,3}\vert =
  r_t \widetilde{m}_{t_1}^2$.
The choice $r_t = r_b$ gives a $\widetilde{t}$-squark sector closely
 mirroring the $\widetilde{b}$-squark sector, and it helps suppressing 
 the production cross section of same-chirality sleptons through a 
 partial cancellation of the D-term contributions.
By pushing all other squark masses at the TeV level, all other
 contributions from D-term couplings are suppressed.

We therefore examine the following squark spectrum:
\begin{equation}
\begin{array}{ll}
&
\begin{array}{l}
  \,(\widetilde{m}_{Q}^2)_{ii}  =
  \,(\widetilde{m}_{{D}}^2)_{ii}=(\widetilde{m}_{{U}}^2)_{ii}
   \, = (1\, {\rm TeV})^2   \quad \quad (i=1,2),
\\[1.1ex]
  \vert (\widetilde{m}^2_{U,{\rm RL}})_{ii}\vert, \, 
  \vert (\widetilde{m}^2_{D,{\rm RL}})_{ii}\vert \ll
  \,(\widetilde{m}_{Q})_{ii}    \quad \qquad \ (i=1,2),
\\[1.1ex]
  \vert (\widetilde{m}^2_{D,{\rm RL}})_{3,3}\vert  = 
      r_b \widetilde{m}_{b_1}^2,  \,
  \vert (\widetilde{m}^2_{U,{\rm RL}})_{3,3}\vert = 
      r_b \widetilde{m}_{t_1}^2,   
  \quad \theta_b = \theta_t = 45^0,  \quad
\\[1.01ex]
  \ \widetilde{m}_{t_1}^2   = \,  
  \left[ \widetilde{m}_{b_1}^2(1+r_b)+m_t^2-(\frac{3}{4})M_Z^2\right]/
  [1+r_b] 
\\[1.1ex]
  \, \widetilde{m}_{b_2}^2   =  (1+2 r_b) \widetilde{m}_{b_1}^2,
  \, \widetilde{m}_{t_2}^2   =  (1+2 r_b) \widetilde{m}_{t_1}^2, 
\end{array}
\label{eq:spectrBsq}
\end{array}
\end{equation}

If we let $\widetilde{m}_{b_1}$ take values in the same interval
 $[100,200]\,$GeV chosen before for the lightest squark masses in the 
 spectrum~{\bf A}, the values of $\widetilde{m}_{b_2}$ span an interval
 fixed by $r_b$, such as $[150,300]\,$GeV for $r_b=62.5\%$.
This choice brings
  $\vert (\widetilde{m}^2_{D,{\rm RL}})_{3,3} \vert^{1/2}$ to be within
 $\sim 80$ and $160\,$GeV, the two $\widetilde{t}$ squark eigenstates 
 $\widetilde{t}_1$ and $\widetilde{t}_2$ to have masses in
 the intervals $\sim [150,230]\,$GeV and $\sim [230,340]\,$GeV, 
 respectively, and 
 $\vert (\widetilde{m}^2_{U,{\rm RL}})_{3,3}\vert^{1/2}$ to lie between
 $\sim 120$ and $180\,$GeV.

\begin{figure}[t] 
\begin{center} 
\epsfxsize= 12.5cm 
\leavevmode 
\epsfbox{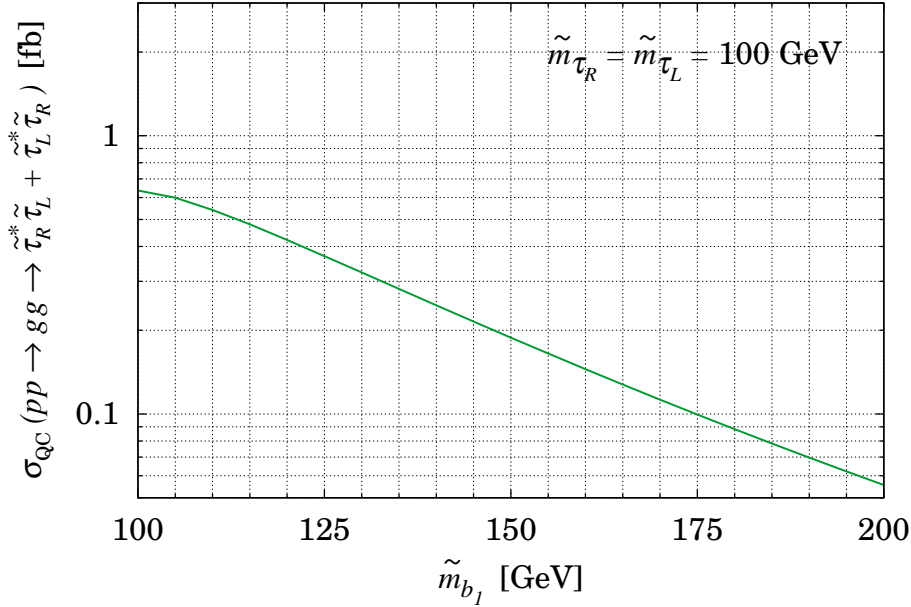} 
\vspace*{-0.5truecm}
\end{center} 
\caption{\small{\it Cross section for the production of a pair of
 $100\,$GeV opposite chirality $\widetilde{\tau}$'s induced by  
 F-term quartic-coupling, for $\tan \beta =30$ and the spectrum
 of Eq.~(\ref{eq:spectrBsq}).
 This may be regarded as an upper bound of the F-term contribution 
 to the cross section; see discussion in the text.}}
\label{figPLOT:QCxscB} 
\end{figure}

We show in Fig.~\ref{figPLOT:QCxscB} the cross section for the
 production of opposite chirality $\widetilde{\tau}$ states of
 $100\,$GeV, assumed here to be mass eigenstates, obtained with this
 spectrum and $\tan\beta=30$.
The cross section decreases from about 0.6$\,$fb to 0.06$\,$fb in this 
 $\widetilde{b}_1$ mass range.
Despite having a slightly steeper shape, it is comparable to the
 cross section for the production of a pair of left-handed sleptons
 shown in Fig.~\ref{figPLOT:QCxscA}.

The values shown in this figure, however, may not be achieved for
 realistic sfermion spectra.
The most serious problem is the magnitude of $\vert(A_D)_{33}\vert$,
 which tends to lie at the $1\,$TeV level or above for all values of
 $r_b$ , when $\tan \beta$ has the value chosen here. 
Such values of $\vert (A_D)_{33}\vert$ lead to a charge beaking
 vacuum~\cite{CHH}, with lifetime shorter than the age of the
 Universe~\cite{LargeAtunn}. 
Towards the higher end of the values of $\widetilde{m}_{b_1}$ in 
 Fig.~\ref{figPLOT:QCxscB}, however, the value of
 $\vert (A_D)_{33}\vert$ can become sufficiently small to guarantee a
 long enough lifetime of the metastable electroweak vacuum,
 provided $\tan \beta$ does not significantly exceed the value of 10. 
An inspection of Fig.~\ref{figPLOT:QCxscB} shows that such a decrease
 in $\tan \beta$ would immediately plunge the corresponding cross
 section to values of ${\cal O}(10^{-2})\,$fb.
An additional suppression comes also from relaxing the assumption
 of vanishing chirality-mixing terms in the $\widetilde{\tau}$-slepton
 sector, made for this figure.
This assumption is also a problematic one, since it induces
 unacceptable values for $\vert (A_E)_{33}\vert$.

We therefore abondon the idea of studying further the contribution from 
 F-term quartic couplings to the slepton production cross section from
 gluon fusion.  
Hereafter, the contribution from quartic couplings to the cross section 
 should be understood as induced by D-terms only.

\section{The one-loop $\mbox{\boldmath{$gg \to \widetilde{l}^{\ast}
                           \widetilde{l}$}}$ cross section}
\label{sec:CompleteXsec}
\begin{figure}[t] 
\begin{center} 
\epsfxsize= 5.31 cm 
\leavevmode 
\epsfbox{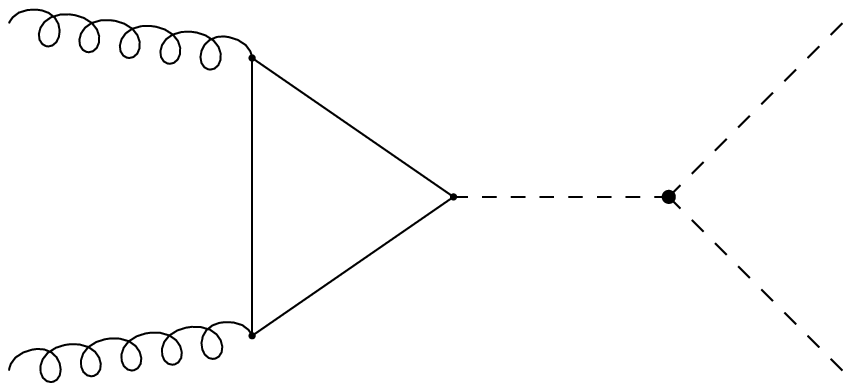} 
\put(-103,26){$q_n$}
\put(-63,18){${\cal H}, A$}
\put(-3,43){$\widetilde{l}^\ast_i$}
\put(-3,6){$\widetilde{l}_j$}
\hspace*{1.3truecm} 
\epsfxsize= 5.4 cm 
\epsfbox{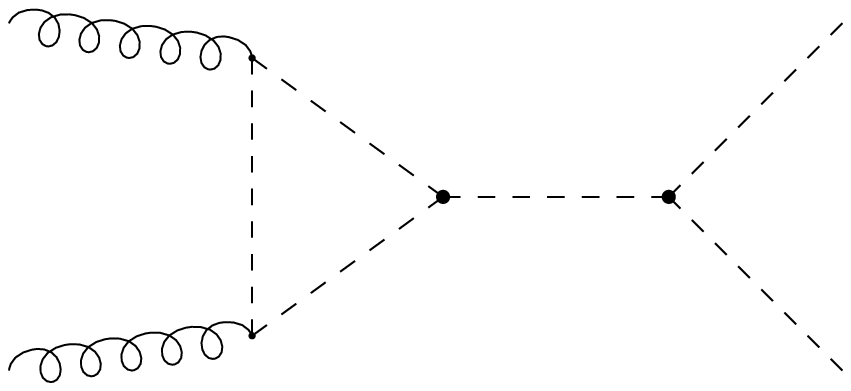} 
\put(-103,26){$\widetilde{q}_n$}
\put(-57,19){${\cal H}$}
\put(-3,43){$\widetilde{l}^\ast_i$}
\put(-3,6){$\widetilde{l}_j$}
\end{center} 
\caption{\small{\it Diagrams contributing to the effective couplings
 $gg\,\widetilde{l}_i^\ast\widetilde{l}_j$, through exchange of a neutral
 Higgs boson in the s-channel. The symbol ${\cal H}$ denotes $h$ and $H$.
}} 
\label{figDIAG:Hmed} 
\end{figure}

%
Effective $g g\,\widetilde{l}_i^\ast \widetilde{l}_j$ interactions 
 can also be induced by the decay into a slepton pair of a neutral Higgs
 boson, produced at the one-loop level through gluon fusion. 
Examples of Feynmann diagrams giving rise to these effective interactions
 are shown in Fig.~\ref{figDIAG:Hmed}. 
There exists also a squark-bubble diagram analogous to that in 
 Fig.~\ref{figDIAG:QCloop}, with a neutral Higgs boson emerging from 
 the bubble to produce the slepton pair.  
The complete cross section including also the contribution from these
 diagrams has a more complicated dependence on the supersymmetric
 parameter space than that obtained only from slepton-squark quartic 
 couplings, as it depends also on the Higgs-sector parameters.

The neutral Higgs boson exchanged in the s-channel is one of the mass
 eigenstates obtained from suitable rotations of the neutral components
 of the two Higgs doublets $H_d^0$ and $H_u^0$, the two scalar Higgs
 bosons, $h$ and $H$, and the pseudoscalar one, $A$:
\begin{eqnarray}
 H_d^0  & = &  \frac{1}{\sqrt{2}}
 \left\{
 v\cos \beta + \left(\cos \alpha H -\sin \alpha h  \right)
             +i\left(\sin \beta  A +\cos \beta \chi^0\right)\right\},
\nonumber \\
 H_u^0  & = &  \frac{1}{\sqrt{2}}
 \left\{
 v\sin \beta + \left(\sin \alpha H +\cos \alpha h  \right)
             +i\left(\cos \beta  A -\sin \beta \chi^0\right)\right\},
\label{eq:HIGGSmassEIG}
\end{eqnarray}
 where $\chi^0$ is the Goldstone boson. 
Once radiative corrections to the Higgs sector are included, the angle
 $\alpha$ is determined by the relation~\cite{SPIRAhabil}:
\begin{equation}
 \tan 2 \alpha =
 \frac{m_A^2 +M_Z^2}{m_A^2-M_Z^2 +\epsilon/\cos2\beta} \tan 2 \beta,
\qquad \quad 
 0 > \alpha > -\frac{\pi}{2}, 
\label{eq:alphaANGLE}
\end{equation}
 where $\epsilon$ is a correction factor proportional to the fourth
 power of the $t$-quark mass: 
\begin{equation}
 \epsilon =
 \frac{3 G_F}{\sqrt{2}\pi^2} \frac{m_t^4}{\sin^2\beta}
 \log\left(1+\frac{\widetilde{m}^2}{m_t^2}\right)
\label{eq:epsilon}
\end{equation}
 with $\widetilde{m}$, the scale of the $\widetilde{t}$ scalar masses.
The mass of the two physical states $h$ and $H$ are
\begin{eqnarray}
 m_h^2  &\leq &  M_Z^2\vert\cos 2\beta \vert +\epsilon \sin^2\beta,
\nonumber \\
 m_H^2  &  =  &  m_A^2 +M_Z^2 -m_h^2 +\epsilon,  
\label{eq:HiggsMASSrel}
\end{eqnarray}
 with the complete expression for $m_h$ given by:
\begin{equation}
 m_h^2 = 
 \frac{m_A^2 \!+\!M_Z^2\!+\!\epsilon}{2} -\! 
 \left[\! 
  \left(\!\frac{m_A^2 \!+\!M_Z^2\!+\!\epsilon}{2}\!\right)^{\!2}
  \!\!-\! m_A^2 M_Z^2 \cos^2\!2\beta 
  -\!\epsilon\left(\!m_A^2 \sin^2\!\beta +\!M_Z^2\cos^2\!\beta\right)
 \right]^{1/2}\!\!\!\!\!\! ,
\label{eq:hMASScompl}
\end{equation}
 whereas $m_A$ can be considered a free parameter.
Although the state-of-the-art of radiative corrections to the Higgs
 sector~\cite{ABDELD} is far more sophisticated than the sketchy
 picture given here, this description is sufficient for our discussion.
In the following numerical evaluations, we use the Higgs boson masses
 and widths from CPSUPERH~\cite{CPSUPERH}.

The three neutral Higgs bosons $h$, $H$, and $A$ couple to a fermion
 $f$ as:
\begin{equation}
  V_{\rm H-ferm} \ = \
    Y_{hff}\bar{f} f h        +   Y_{Hff} \bar{f}f H 
+i\ Y_{Aff}\bar{f}\gamma_5 f A+i\ Y_{\chi^0 ff}\bar{f}\gamma_5 f\chi^0,
\label{eq:Hfermions}
\end{equation}
 with 
\begin{eqnarray}
 Y_{htt}\  =\   \frac{\left(h_U\right)_{3,3}}{\sqrt{2}}\cos\alpha, 
&&
 Y_{Htt}   =\ \ \frac{\left(h_U\right)_{3,3}}{\sqrt{2}}\sin\alpha,
\nonumber \\
 Y_{Att}\, =\   \frac{\left(h_U\right)_{3,3}}{\sqrt{2}}\cos\beta,
&&
 Y_{\chi^0tt}=- \frac{\left(h_U\right)_{3,3}}{\sqrt{2}}\sin\beta,  
\label{eq:yttDEFs}
\end{eqnarray}
 when $f$ is the $t$ quark, and 
\begin{eqnarray}
 Y_{hbb}\ =- \frac{\left(h_D\right)_{3,3}}{\sqrt{2}} \sin\alpha, 
&&
 Y_{Hbb}  =\ \frac{\left(h_D\right)_{3,3}}{\sqrt{2}} \cos\alpha,
\nonumber \\
 Y_{Abb}\,=\ \ \frac{\left(h_D\right)_{3,3}}{\sqrt{2}} \sin\beta,
&&
 Y_{\chi^0bb}=\ \frac{\left(h_D\right)_{3,3}}{\sqrt{2}} \cos\beta,
\label{eq:ybbDEFs}
\end{eqnarray}
 when it is the $b$ quark.

\begin{table}[p]
\begin{center} 
\begin{tabular}{cccccc} 
\hline\hline                                       
&&                                                 &
\\[-1.9ex]       
Higgs(es)                                          & 
$s_ms_n$                                            & 
$\qquad d^{H_i^0 H_i^0\!,\,\widetilde{s}_n\widetilde{s}_n}$       & 
$\quad f^{H_i^0 H_i^0\!,\,\widetilde{s}_n\widetilde{s}_n}$        & 
$\, f_\mu^{H_i^{0\ast}\!\!,\,\widetilde{s}_m\widetilde{s}_n}$       &
$\, t^{H_i^0\!,\widetilde{s}_m\widetilde{s}_n}\quad$            
\\[1.01ex]
\hline                                                
&&                                                &
\\[-1.9ex]     
$\quad H_d^{0\,\ast} H_d^0 \quad \quad$               & 
$\widetilde{u}_L^\ast\widetilde{u}_L$               & 
$\quad\displaystyle{+\frac{g_2^2}{4}} -\frac{g_1^2}{12}$ & 
$-$                                                & 
$-$                                                & 
$-$                                                      
\\[1.4ex] 
$\quad H_d^{0\,\ast} H_d^0 \quad \quad$                & 
$\widetilde{d}_L^\ast\widetilde{d}_L$                & 
$\quad\displaystyle{-\frac{g_2^2}{4}} -\frac{g_1^2}{12}$ & 
$ h_D^2$                                            &
$-$                                                & 
$-$                                                      
\\[1.4ex] 
$\quad H_d^{0\,\ast} H_d^0 \quad \quad$                & 
$\widetilde{u}_R^\ast\widetilde{u}_R$                & 
$\quad+\displaystyle{\frac{g_1^2}{3}}$              & 
$-$                                                & 
$-$                                                & 
$-$                                                      
\\[1.5ex] 
$\quad H_d^{0\,\ast} H_d^0 \quad \quad$                & 
$\widetilde{d}_R^\ast\widetilde{d}_R$                & 
$\quad -\displaystyle{\frac{g_1^2}{6}}$             & 
$ h_D^2$                                            &
$-$                                                & 
$-$                                                      
\\[1.4ex] 
$\quad H_d^{0\,\ast} \quad \quad$                     & 
$\widetilde{t}_R^\ast\widetilde{t}_L$                & 
$\qquad-$                                          & 
$-$                                                &
$-\mu (h_U)_{3,3}$                                   & 
$-$                                                      
\\[1.4ex] 
$\quad H_d^0 \, \quad \quad$                        & 
$\widetilde{b}_R^\ast\widetilde{b}_L$                & 
$\qquad-$                                          & 
$-$                                                &
$-$                                                & 
$({A}_D)_{3,3}(h_D)_{3,3}$                                
\\[1.01ex] 
\hline                                                
&&                                                 &
\\[-1.9ex]     
$\quad H_d^{0\,\ast} H_d^0 \quad \quad$                & 
$\widetilde{l}_L^\ast\widetilde{l}_L$                & 
$\quad\displaystyle{-\frac{g_2^2}{4}} +\frac{g_1^2}{4}$ & 
$ h_E^2$                                            &
$-$                                                & 
$-$                                                      
\\[1.4ex] 
$\quad H_d^{0\,\ast} H_d^0 \quad \quad$                & 
$\widetilde{l}_R^\ast\widetilde{l}_R$                & 
$\quad -\displaystyle{\frac{g_1^2}{2}}$             & 
$ h_E^2$                                            &
$-$                                                & 
$-$                                                      
\\[1.5ex] 
$\quad H_d^0 \, \quad \quad$                        & 
$\widetilde{\tau}_R^\ast\widetilde{\tau}_L$          & 
$\qquad-$                                          & 
$-$                                                &
$-$                                                & 
$({A}_E)_{3,3} (h_E)_{3,3}$                               
\\[1.4ex] 
\hline                                          
&&                                                 &
\\[-1.9ex]        
\hline                                          
&&                                                 &
\\[-1.9ex]       
$\quad H_u^{0\,\ast} H_u^0 \quad \quad$                & 
$\widetilde{u}_L^\ast\widetilde{u}_L$                & 
$\quad\displaystyle{-\frac{g_2^2}{4}} +\frac{g_1^2}{12}$ & 
$ h_U^2$                                            & 
$-$                                                & 
$-$                                                      
\\[1.4ex] 
$\quad H_u^{0\,\ast} H_u^0 \quad \quad$                & 
$\widetilde{d}_L^\ast\widetilde{d}_L$                & 
$\quad\displaystyle{+\frac{g_2^2}{4}} +\frac{g_1^2}{12}$ & 
$-$                                                &
$-$                                                & 
$-$                                                      
\\[1.4ex] 
$\quad H_u^{0\,\ast} H_u^0 \quad \quad$                & 
$\widetilde{u}_R^\ast\widetilde{u}_R$                & 
$\quad-\displaystyle{\frac{g_1^2}{3}}$              & 
$h_U^2$                                             & 
$-$                                                & 
$-$                                                      
\\[1.5ex] 
$\quad H_u^{0\,\ast} H_u^0 \quad \quad$                & 
$\widetilde{d}_R^\ast\widetilde{d}_R$                & 
$\quad +\displaystyle{\frac{g_1^2}{6}}$             & 
$-$                                                &
$-$                                                & 
$-$                                                      
\\[1.4ex] 
$\quad H_u^0\, \quad \quad$                         & 
$\widetilde{t}_R^\ast\widetilde{t}_L$                & 
$\qquad-$                                          & 
$-$                                                &
$-$                                                & 
$({A}_U)_{3,3} (h_U)_{3,3}$                               
\\[1.4ex] 
$\quad H_u^{0\,\ast} \, \quad \quad$                  & 
$\widetilde{b}_R^\ast\widetilde{b}_L$                & 
$\qquad-$                                          & 
$-$                                                &
$-\mu (h_D)_{3,3}$                                   & 
$-$                                                      
\\[1.01ex] 
\hline                                                
&&                                                 &
\\[-1.9ex]     
$\quad H_u^{0\,\ast} H_u^0 \quad \quad$                & 
$\widetilde{l}_L^\ast\widetilde{l}_L$                & 
$\quad\displaystyle{+\frac{g_2^2}{4}} -\frac{g_1^2}{4}$  & 
$-$                                                &
$-$                                                & 
$-$                                                      
\\[1.4ex] 
$\quad H_u^{0\,\ast} H_u^0 \quad \quad$                & 
$\widetilde{l}_R^\ast\widetilde{l}_R$                & 
$\quad +\displaystyle{\frac{g_1^2}{2}}$             & 
$-$                                                &
$-$                                                & 
$-$                                                      
\\[1.5ex] 
$\quad H_u^{0\,\ast} \quad \quad$                     & 
$\widetilde{\tau}_R^\ast\widetilde{\tau}_L$          & 
$\qquad-$                                          & 
$-$                                                &
$-\mu (h_E)_{3,3}$                                   & 
$-$                                                      
\\[1.4ex] 
\hline\hline
\end{tabular}
\caption{\small{\it Scalar potential couplings of the interactions among
 the neutral components of the two Higgs doublets and sfermions.
 The Hermitian conjugates of the interactions listed here between one
 Higgs boson and two opposite-chirality sfermions also exist.
 In the approximation of real SUSY-breaking parameters and real $\mu$ 
 their couplings coincide with those given in this table.
}}
\label{table:GEHiggsSCALcoupls}
\end{center}
\end{table}

The trilinear couplings of $h$, $H$, and $A$ to squarks and sleptons
 originate from the trilinear soft SUSY-breaking terms, which can be 
 read off from Eqs.~(\ref{eq:trilinear}) and~(\ref{eq:LRmixingMass})
 when $v_i/\sqrt{2}$ is replaced by $H_i$, $i=u,d$, and from the
 superpotential $\mu$ term:
\begin{equation}
 V_{\rm tril} \ = \
 t^{H_i^0\!,\widetilde{s}_R\widetilde{s}_L} \,
 H_i^0        (\widetilde{s}_R^\ast \widetilde{s}_L)  + 
 f_\mu^{H_i^{0\ast}\!,\widetilde{s}_R\widetilde{s}_L}\,
 H_i^{0\,\ast} (\widetilde{s}_R^\ast \widetilde{s}_L)  +{\rm H.c.} 
\qquad 
\left\{\begin{array}{l} s=u,d,l \\ i = u,d \end{array} \right. ,
\label{eq:Vtril}
\end{equation}
 as well as from quartic couplings from D- and F-terms:
\begin{equation}
 \begin {array}{lll}
  V_D & \supset & 
     d^{H_i^0 H_i^0\!,\widetilde{s}_n \widetilde{s}_n} \,
     (H_i^{0\,\ast} H_i^0) (\widetilde{s}_n^\ast \widetilde{s}_n) ,
  \\[1.01ex]
  V_F & \supset &
     f^{H_i^0 H_i^0\!,\widetilde{s}_n \widetilde{s}_n} \,
     (H_i^{0\,\ast} H_i^0) (\widetilde{s}_n^\ast \widetilde{s}_n) , 
  \end{array}
\qquad
 \left\{
  \begin{array}{l} 
     s_n = u_L, d_L, l_L, u_R, d_R, l_R  \\ i = u,d 
  \end{array} \right. ,
\label{eq:VDVFsfermHiggs}
\end{equation}
 in which one of the two neutral Higgs bosons acquires a vacuum
 expectation value.
A list of these couplings is given in
 Table~\ref{table:GEHiggsSCALcoupls}.

In terms of the neutral mass eigenstate Higgs bosons, these trilinear
 interactions have the compact form:
\begin{eqnarray}
 V_{\rm H-sfermion} & \supset & 
      v \left(
  g_{\,h \widetilde{s}_n \widetilde{s}_n} h +
  g_{H \widetilde{s}_n \widetilde{s}_n} H \right)
                  (\widetilde{s}_n^\ast \widetilde{s}_n) + 
\nonumber \\     &&
  \left[v \left(
  g_{\,h \widetilde{s}_R \widetilde{s}_L} h +  
  g_{H \widetilde{s}_R \widetilde{s}_L} H + i\ 
  g_{A \widetilde{s}_R \widetilde{s}_L} A +i\ 
  g_{\chi^0 \widetilde{s}_R \widetilde{s}_L} \right) 
                  (\widetilde{s}^\ast_R \widetilde{s}_L) + {\rm H.c.} 
 \right],
\label{eq:potHIGGSsferm}
\end{eqnarray}
 with couplings listed explicitly in Table~\ref{table:MEHiggsSCALcoupls}.

\begin{table}[t]
\begin{center} 
\begin{tabular}{ccc} 
\hline\hline                                           
&&
\\[-1.9ex]       
                                                    & 
$s_n = l_L, l_R \,(\!h_S\!=\!h_E\!);
       d_L, d_R \,(\!h_S\!=\! h_D\!)$                 &
$s_n = u_L, u_R $                                     
\\[1.01ex]    
\hline                                               
&&
\\[-1.9ex]     
$g_{\,h \widetilde{s}_n \widetilde{s}_n}$                        & 
$\left[ s_{(\beta+\alpha)} \,d^{H_u^0 H_u^0\!,s_n s_n}
        - h_S^2 \,c_\beta s_\alpha \right]$              &  
$\left[ s_{(\beta+\alpha)} \,d^{H_u^0 H_u^0\!,s_n s_n}
        + h_U^2 \,s_\beta c_\alpha \right]$      
\\[1.4ex]    
$g_{H \widetilde{s}_n \widetilde{s}_n}$                         &
$\left[ c_{(\beta+\alpha)} \,d^{H_d^0 H_d^0\!,s_n s_n}
        + h_S^2 \,c_\beta c_\alpha \right]$              & 
$\left[ c_{(\beta+\alpha)} \,d^{H_d^0 H_d^0\!,s_n s_n}
        + h_U^2 \,s_\beta s_\alpha \right]$              
\\[1.4ex] 
\hline\hline                                           
&&
\\[-1.9ex]       
                                                     & 
$s = l\,(\!h_S\!=\!h_E,\!{A}_S\!=\!{A}_E);
     d\,(\!h_S\!=\!h_D,\!{A}_S\!=\!{A}_D)$            &
$s = u$                                                  
\\[1.01ex]    
\hline                                                 
&&
\\[-1.9ex]     
$g_{\,h \widetilde{s}_R \widetilde{s}_L}$                        & 
$\phantom{i}(-{A}_S\,s_\alpha -\mu c_\alpha)h_S/
            (\sqrt{2}\,v)$                           & 
$\phantom{i}(+{A}_U\,c_\alpha +\mu s_\alpha)h_U/
            (\sqrt{2}\,v)$                               
\\[1.4ex]
$g_{H \widetilde{s}_R \widetilde{s}_L}$                          & 
$\phantom{i}(+{A}_S\,c_\alpha -\mu s_\alpha)h_S/
            (\sqrt{2}\,v)$                           & 
$\phantom{i}(+{A}_U\,s_\alpha -\mu c_\alpha)h_U/
            (\sqrt{2}\,v)$                               
\\[1.4ex]
$g_{A \widetilde{s}_R \widetilde{s}_L}$                          & 
$ (+{A}_S\,s_\beta +\mu c_\beta)h_S/
  (\sqrt{2}\,v)$                                     & 
$ (+{A}_U\,c_\beta +\mu s_\beta)h_U/
  (\sqrt{2}\,v)$                                         
\\[1.4ex]
$g_{\chi^0 \widetilde{s}_R \widetilde{s}_L}$                      & 
$ (+{A}_S\,c_\beta -\mu s_\beta)h_S/
  (\sqrt{2}\,v)$                                     & 
$ (-{A}_U\,s_\beta +\mu c_\beta)h_U/
  (\sqrt{2}\,v)$                                     \\[1.4ex] 
\hline\hline
\end{tabular} 
\caption{\small{\it Couplings of sfermions to the mass eigenstate
 neutral Higgs bosons ${\cal H}=\{h,H\}$ and $A$.
 The couplings
 $g_{{\cal H} \widetilde{s}_L \widetilde{s}_R}$,
 $g_{A \widetilde{s}_L \widetilde{s}_R}$, and
 $g_{\chi^0 \widetilde{s}_L \widetilde{s}_R}$, not explicitly listed, coincide
 respectively with
 $g_{{\cal H} \widetilde{s}_R \widetilde{s}_L}$,
 $g_{A \widetilde{s}_R \widetilde{s}_L}$, and
 $g_{\chi^0 \widetilde{s}_R \widetilde{s}_L}$ in
 the approximation of real SUSY-breaking parameters and real $\mu$.
 The symbols $c_x$, $s_x$ are abbreviations of $\cos x$, $\sin x$.
 Only the third generation Yukawa couplings  $(h_U)_{3,3}$, $(h_D)_{3,3}$,
 and $(h_E)_{3,3}$ are assumed to be nonvanishing.
}}
\label{table:MEHiggsSCALcoupls}
\end{center}
\end{table}

We note here that also in this case sum rules analogous to those in
 Eq.~(\ref{eq:chSUMRULE1}) exist for each generation of squarks:
\begin{equation}
 \sum_{\widetilde{q}_n}
 d^{H_u^0 H_u^0\!,\,\widetilde{q}_n\widetilde{q}_n} \ =\ 
 \sum_{\widetilde{q}_n}
 d^{H_d^0 H_d^0\!,\,\widetilde{q}_n\widetilde{q}_n} \ = 0.
\label{eq:chSUMRULE2}
\end{equation}
As a consequence, similar sum rules hold for the D-terms contributions
 to the couplings $g_{\,h \widetilde{q}_n \widetilde{q}_n}$ and
 $g_{H \widetilde{q}_n \widetilde{q}_n}$.

This mechanism of production of a pair of sleptons through gluon
 fusion may also be regarded as a test of SUSY, although triple couplings
 of partners of the SM particles are not unique of SUSY models.
In particular, unlike the Drell-Yan production mechanism, 
 the gluon-fusion mechanism mediated by the exchange of a Higgs boson
 can probe the Higgs sector, the $\mu$ parameter, $\tan \beta$, and the
 trilinear $A$ terms.

The parton-level cross section for the gluon-fusion production of a pair
 of sleptons, can be expressed in terms of two form factors,
 which parametrize the effective Lagrangian induced at the one-loop
 level by quartic-couplings and Higgs-exchange diagrams:
\begin{equation}
 {\cal L}_{gg \to \widetilde{l}_i^\ast \widetilde{l}_j} =
 \frac{\alpha_s}{4 \pi} 
 \left\{
 A^{\,\widetilde{l}_i \widetilde{l}_j}_S 
  \left(g^{\mu \nu}\! -\frac{2 k_2^\mu k_1^\nu}{\hat{s}}\right)
+A^{\,\widetilde{l}_i \widetilde{l}_j}_P   \epsilon^{\mu\nu}_{\alpha\beta}
  \left(\frac{2 k_1 k_2}{\hat{s}}\right)
 \right\}
 \delta_{ab} \,\epsilon^\alpha_\mu(k_1) \epsilon^\beta_\nu(k_2)\, 
 \widetilde{l}_i^\ast \widetilde{l}_j.  
\label{eq:effLAGR}
\end{equation}
Here $\epsilon_{\mu\nu\alpha\beta}$ is a fully antisymmetric tensor,
 with $\epsilon_{0123} =1$, $k_1$ and $k_2$ are the momenta of the two
 initial gluons, and $A^{\,l_i l_j}_P$ is generated by the diagram with
 exchange of the pseudoscalar Higgs boson $A$, whereas all the other
 contributions are included in $A^{\,\widetilde{l}_i \widetilde{l}_j}_S$:
\begin{equation}
  A^{\,\widetilde{l}_i \widetilde{l}_j}_S =
\widetilde{A}_{\,\rm QC}^{\,\widetilde{l}_i \widetilde{l}_j} + 
          {A}_{S,\,{\cal H}}^{\,\widetilde{l}_i \widetilde{l}_j}+ 
\widetilde{A}_{S,\,{\cal H}}^{\,\widetilde{l}_i \widetilde{l}_j}. 
\label{eq:AS}
\end{equation}
In this expression, the tilde distinguishes the contributions from
 scalar loops
 ($\widetilde{A}_{\,\rm QC}^{\,\widetilde{l}_i \widetilde{l}_j}$ is the
 truncated amplitude discussed in the previous section), and the
 subscript \{${\cal H}$\} refers to diagrams with exchange of a
 scalar Higgs boson, {\it i.e.} ${\cal H}=h,H$. 
Notice that, because of our assumption of imaginary
 $i g_{A \widetilde{s}_L \widetilde{s}_R}$
 couplings and real chirality-mixing terms in the sfermion sector, there
 is no contribution to $ A^{\,\widetilde{l}_i \widetilde{l}_j}_S$ from
 the exchange of the pseudoscalar Higgs boson.

In terms of
 $A^{\,\widetilde{l}_i \widetilde{l}_j}_S$ and
 $A^{\,\widetilde{l}_i \widetilde{l}_j}_P$ the parton-level cross
 section is
\begin{equation}
 \widehat{\sigma}(gg\to \widetilde{l}_i^\ast \widetilde{l}_j) =  
 \frac{\alpha_s^2}{(16\pi)^3} \, 
 \frac{\lambda^{1/2}(\hat{s},\widetilde{m}_{l_i},\widetilde{m}_{l_j})}
      {\hat{s}} \,
 \left\{ \left\vert {A}^{\,\widetilde{l}_i \widetilde{l}_j}_S
         \right\vert^2 +
         \left\vert {A}^{\,\widetilde{l}_i \widetilde{l}_j}_P
         \right\vert^2 \right\} .
\label{eq:totalXsec}
\end{equation}

The expression for
 ${A}_{S,\,{\cal H}}^{\,\widetilde{l}_i \widetilde{l}_j}$ and
 $A^{\,\widetilde{l}_i \widetilde{l}_j}_P$ (which strictly speaking
 should be labelled
 $A^{\,\widetilde{l}_i \widetilde{l}_j}_{P,\,A}$) coming from
 fermion loops are
\begin{eqnarray}
 {A}_{S,\,{\cal H}}^{\,\widetilde{l}_i \widetilde{l}_i} \!&=& \!\!
  - \!\! \sum_{{\cal H}=h,H} \, 
  g_{{\cal H} \widetilde{l}_i \widetilde{l}_j} \left(\frac{v}{m_t}\right)
 \left(
 \frac{1}{1-\!  m_{\cal H}^2/\hat{s}
           +\!i m_{\cal H}\Gamma_{\cal H}/\hat{s} }
 \right)
  Y_{{\cal H}tt} \,F(\hat{s};m_t^2) , 
\label{eq:ASfermion}
\\[1.001ex]
 A^{\,\widetilde{l}_L \widetilde{l}_R}_P           \!&=& \!\! 
  \quad \, -\quad
  g_{{\cal A} \widetilde{l}_L \widetilde{l}_R}
  \left(\frac{v}{m_t}\right)
  \left(
  \frac{1}{1-\!  m_A^2/\hat{s} +\!i m_A\Gamma_A/\hat{s}} \right)
   Y_{Att} \,P(\hat{s};m_t^2) ,
\label{eq:APfermion}
\end{eqnarray}
 with
 $A^{\,\widetilde{l}_L \widetilde{l}_R}_P =
  A^{\,\widetilde{l}_R \widetilde{l}_L}_P$ and
 $A^{\,\widetilde{l}_L \widetilde{l}_L}_P =
  A^{\,\widetilde{l}_R \widetilde{l}_R}_P=0$, 
 that for 
 $\widetilde{A}_{S,\,{\cal H}}^{\,\widetilde{l}_i \widetilde{l}_i}$
 is
$$
\ \widetilde{A}_{S,\,{\cal H}}^{\,\widetilde{l}_i \widetilde{l}_i} \ = \
   -\!\!\sum_{{\cal H}=h,H} \,
   g_{{\cal H} \widetilde{l}_i \widetilde{l}_i} 
  \left(\frac{v^2}{\hat{s}}\right)
  \left(
 \frac{1}{1 -\!m_{\cal H}^2/\hat{s}
            +\!i m_{\cal H}\Gamma_{\cal H}/\hat{s} }
  \right) \times \hspace{2.8truecm} \vspace*{-0.7truecm}
$$
\begin{eqnarray}
 \hspace*{2.3truecm}
 \sum_{q =u,d}  \hspace*{-0.5truecm}
&& 
\left[\left( c_q^2\, 
 g_{{\cal H}\widetilde{q}_L\widetilde{q}_L}
            +s_q^2\,
 g_{{\cal H}\widetilde{q}_R\widetilde{q}_R}
         +2 (s_q c_q)
 g_{{\cal H}\widetilde{q}_R\widetilde{q}_L}
      \right)
       S(\hat{s};\widetilde{m}_{q_1}^2) \,+
\right.
\nonumber\\[-1ex] 
&& \left. \
\left( s_q^2\,
  g_{{\cal H}\widetilde{q}_L\widetilde{q}_L}
      +c_q^2\,
  g_{{\cal H}\widetilde{q}_R\widetilde{q}_R}
        -2 (s_q c_q)
  g_{{\cal H}\widetilde{q}_R\widetilde{q}_L} \right)
     S(\hat{s};\widetilde{m}_{q_2}^2)
  \right],  
\label{eq:ASsfermion}
\end{eqnarray}
 with $c_q=\cos q$ and $s_q = \sin q$, and $q$ spanning over all three
 generations.   
We remind that in the limit of vanishing chirality-mixing terms in the
 squark sector, $\cos q \to 0$, $\sin q \to 1$, and 
 $\widetilde{q}_1 \to \widetilde{q}_R$,
 $\widetilde{q}_2 \to \widetilde{q}_L$.

\begin{figure}[t] 
\begin{center} 
\epsfxsize= 12.5cm 
\leavevmode 
\epsfbox{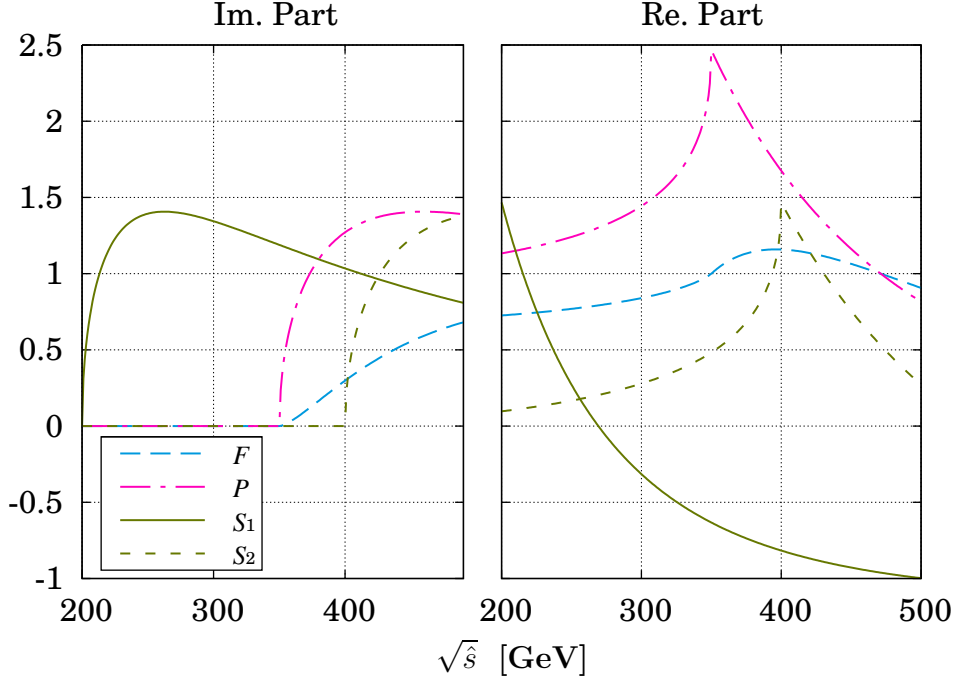} 
\vspace*{-0.5truecm}
\end{center} 
\caption{\small{\it Imaginary and real part of the functions
  $F(\hat{s};m^2)$, $P(\hat{s};m^2)$, and $S(\hat{s};m^2)$ versus
  $\hat{s}$. 
 The value of $m$ is fixed to be $m_t$ for $F$ and $P$, shown by the
  azure long-dashed line and the magenta dot-dashed one, respectively.
 Two values of $m$ are chosen for $S$, shown by the two olive lines:
  $m=100\,$GeV~(solid line $S_1$), and  $m=200\,$GeV~(short-dashed
 line $S_2$).
}} 
\label{fig:SFPfuncs} 
\end{figure}

The function $S(\hat{s};m_{\widetilde{q}_n}^2)$ is given in
 Eq.~(\ref{eq:functionS}) in Sec.~\ref{sec:QCXsecANAL},
 $F(\hat{s};m_t^2)$ and $P(\hat{s};m_t^2)$ are 
\begin{eqnarray}
 F(\hat{s};m^2) &=& \tau \left[1 +(1-\tau) f(\tau)\right],  
\nonumber\\
 P(\hat{s};m^2) &=& \tau \,f(\tau),   
 \hspace*{3.5truecm}  
 \tau \equiv 4 m^2/\hat{s},
\label{eq:functionF} 
\end{eqnarray}
 with $f(\tau)$ already defined in Eq.~(\ref{eq:functionf}). 
The imaginary and real parts of the three functions $F(\hat{s};m^2)$,
 $P(\hat{s};m^2)$, and $S(\hat{s};m^2)$ are plotted in
 Fig.~\ref{fig:SFPfuncs} versus $\hat{s}$.
We have taken $m=m_t$ for $F$ and $P$, and the two values $m=100$ and
 $200\,$GeV for $S$.
The two resulting functions of $\hat{s}$ in this last case are called
 $S_1$ and $S_2$ in this figure. 
See the two olive lines, solid for $S_1$ and short-dashed for $S_2$.
The azure long-dashed line and the red dot-dashed one represent $F$ and
 $P$, respectively.

It should be noted here that, because of the sum rules in 
 Eqs.~(\ref{eq:chSUMRULE1}) and~(\ref{eq:chSUMRULE2}), the constant 
 term -1 in the definition of the function $S$ does not contribute to
 the evaluation of 
 $\widetilde{A}_{\,\rm QC}^{\,\widetilde{l}_i \widetilde{l}_j}$ and 
 $\widetilde{A}_{S,\,{\cal H}}^{\,\widetilde{l}_i \widetilde{l}_j}$, 
 when the quartic and trilinear couplings considered here are 
 determined by D-terms.  
In this case, the function $S$ contributes to these truncated
 amplitudes exactly like the function $P$. 
The values of the relevant real part of $S$ is then obtained shifting
 upward by the constant value 1 the two olive lines in
 Fig.~\ref{fig:SFPfuncs}.

The functions $F$ and $P$ (and $S$, when the relevant couplings come
 from D-terms) vanish in the limit $\hat{s}\to \infty$, for fixed $m$,
 and also in the limit $m \to 0$, for fixed $\hat{s}$.
For $m=m_b$, both the real and imaginary part of $F$ and $P$ are at most
 of ${\cal O}(10^{-3})$.
Nevertheless, the contribution from  a bottom-quark loop, with
 truncated amplitudes analogous to those from the top quark given in 
 Eqs.~(\ref{eq:ASfermion}) and~(\ref{eq:APfermion}), is enhanced by
 the factor $(v/m_b)$, and unsuppressed values of the couplings
 $Y_{Hbb}$ and $Y_{Abb}$ in the case of large $\tan \beta$.
Depending on the value of $\tan \beta$, it may end up being the 
 dominant contribution.

The generalization of the formulae in
 Eqs.~(\ref{eq:ASfermion}),~(\ref{eq:APfermion}),
 and~(\ref{eq:ASsfermion}) to the case with nonvanishing 
 chirality-mixing terms in the slepton sector can be easily obtained
 using the algorithm of Eq.~(\ref{eq:genQCXsecGENslep}) applied to 
 $A_S^{\,l_i l_j}$ and $A_P^{\,l_i l_j}$.

\subsection{Results}
\label{sec:complXsecNUMER}
%
Even before presenting numerical results, it is possible to draw some 
 general observations on the truncated amplitudes induced by the 
 Higgs-boson exchange from an inspection of the couplings
 $Y_{{\cal H}tt}$, $Y_{Att}$, 
 $g_{{\cal H}\widetilde{q}_i \widetilde{q}_j}$, and
 $g_{A \widetilde{q}_i \widetilde{q}_j}$, as well as 
 $g_{{\cal H}\widetilde{l}_i \widetilde{l}_j}$, and
 $g_{A \widetilde{l}_i \widetilde{l}_j}$, for a given Higgs spectrum.

We shall examine below three different Higgs spectra.
Spectrum~{\bf I} has all three Higgs bosons, $h$, $H$, and $A$, light
 and maximally mixed, {\it i.e.}
 $\cos\alpha \sim -\sin\alpha \sim 1/\sqrt{2}$.
All the three states have masses below the threshold
 $\sqrt{\hat{s}_{\rm min}} = 2 \widetilde{m}_{l_i}$. 
Thus, no resonance is encountered when integrating $\hat{s}$ from
 $\hat{s}_{\rm min}$ to $s$. 
Spectrum~{\bf II} has $m_A \gsim 2 \widetilde{m}_{l}$, {\it i.e.}
 $m_A \gsim 200\,$GeV in our case study; $m_h$ and $m_H$ are calculated 
 as explained at the beginning of this section, and depend on
 $\tan \beta$ and the specific squark sector chosen. 
In this case, it is in general $\cos\alpha \sim 1$, and $\sin \alpha$
 is small, but nonnegligible. 
The resonant contributions of $H$ and $A$ are expected to lift up the
 values of our cross sections. 
Spectrum~{\bf III} has $A$ and $H$ rather heavy and completely decoupled
 from our problem. 
The mixing angle $\alpha$ is such that $\cos \alpha = 1$ and
 $\sin \alpha = 0$, to a very good accuracy.

We consider first the case of vanishing chirality-mixing terms in the
 squark sector, as in the squark spectrum~{\bf A} specified in the 
 previous section.
In this context we start concentrating on the couplings
 $g_{{\cal H}\widetilde{q}_L \widetilde{q}_L}$ and
 $g_{{\cal H}\widetilde{q}_R \widetilde{q}_R}$, the only ones giving a nonvanishing
 contribution to
  $\widetilde{A}_{S,\,{\cal H}}^{\,\widetilde{l}_i \widetilde{l}_i}$.

Because of the sum rule in Eq.~(\ref{eq:chSUMRULE2}), also the
 contribution to
 $\widetilde{A}_{S,\,{\cal H}}^{\,\widetilde{l}_i \widetilde{l}_i}$
 from D-terms, as the contribution to 
 $\widetilde{A}_{\,\rm QC}^{\,\widetilde{l}_i \widetilde{l}_i}$,
 vanishes identically in the limit of a completely degenerate
 squark spectrum. 
The squark spectrum~{\bf A}, which we shall use in the following, was
 however conceived precisely with the aim of evading the effectiveness
 of this sum rule.

Unlike in the case of  
 $\widetilde{A}_{\,\rm QC}^{\,\widetilde{l}_i \widetilde{l}_i}$, F-terms give now
 contributions to
 $\widetilde{A}_{S,\,{\cal H}}^{\,\widetilde{l}_i \widetilde{l}_i}$,
 that are nonvanishing.
By inspecting the couplings
 $g_{{\cal H}\widetilde{q}_L \widetilde{q}_L}$ and
 $g_{{\cal H}\widetilde{q}_R \widetilde{q}_R}$ in
 Table~\ref{table:MEHiggsSCALcoupls}, it is easy to see that F-terms
 assign a privileged  role to the $\widetilde{t}$ squarks. 
All other F-term contributions to these couplings are small
 even for large values of $\tan \beta$.
This holds in particular also for the $\widetilde{b}$ squarks,
 whose couplings receive an F-term contribution always multiplied by 
 $\cos\beta$.

In addition, the F-term contributions to the couplings
 $g_{{\cal H}\widetilde{t}_L \widetilde{t}_L}$ and
 $g_{{\cal H}\widetilde{t}_R \widetilde{t}_R}$  are in general larger
 than the contributions from D-terms to any of the couplings
 $g_{{\cal H}\widetilde{q}_L \widetilde{q}_L}$ and
 $g_{{\cal H}\widetilde{q}_R \widetilde{q}_R}$, for any $q$.
In the case of the $h$ exchange, when, however, the cross section is 
 expected to be small, 
 $(h_U\!)_{33}^{\,2}\sin\beta\cos\alpha $ dominates over 
 $\vert \sin(\beta+\!\alpha)
   d^{H_u^0 H_u^0\!,\,\widetilde{t}_R \widetilde{t}_R}\vert$ and 
 $\vert \sin(\beta+\!\alpha)
   d^{H_u^0 H_u^0\!,\,\widetilde{t}_L \widetilde{t}_L}\vert$ by a
 considerable margin.
Similarly, 
 $\vert (h_U\!)_{33}^{\,2}\sin\beta\sin\alpha\vert$ dominates also over  
 $\vert \cos(\beta+\!\alpha)
   d^{H_d^0 H_d^0\!,\,\widetilde{t}_R \widetilde{t}_R}\vert$ and 
 $\vert \cos(\beta+\!\alpha)
   d^{H_d^0 H_d^0\!,\,\widetilde{t}_L \widetilde{t}_L}\vert$, 
 in the case of the heavier Higgs boson exchange $H$, except when 
 $\sin\alpha$ is very close to its typical decoupling-limit value, 
 $\sin\alpha=0$. 
In this limit, however, $H$ is too heavy to be of any relevance for 
 our cross sections.

\begin{figure}[p] 
\begin{center} 
\epsfxsize= 12.5cm 
\leavevmode 
\epsfbox{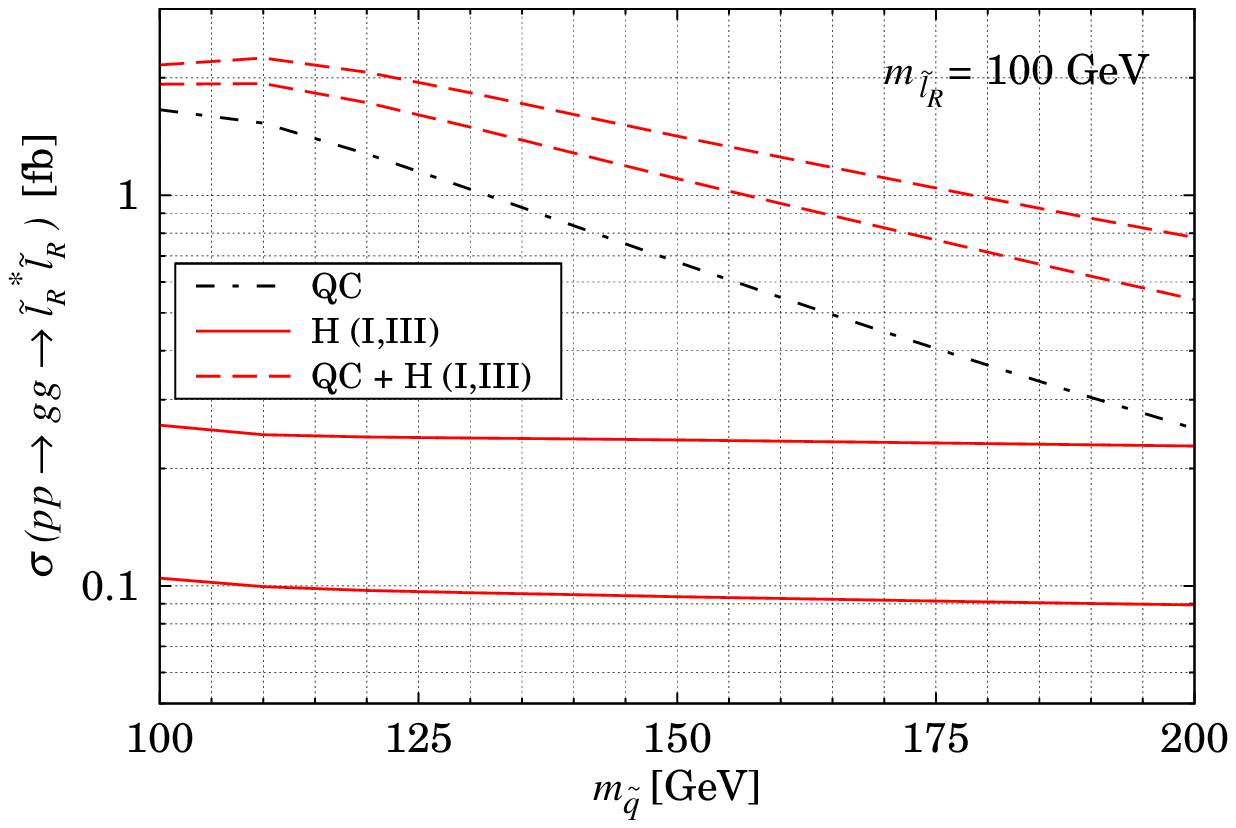} 
\end{center} 
\vspace*{-0.5truecm}
\caption{\small{\it LHC cross section versus the parameter
 $\widetilde{m}_{q}$ in the squark spectrum~{\bf A} for the pair
 production of right-handed sleptons induced by: quartic couplings
 only (black dot-dashed line),
 Higgs exchange only (red solid lines, the upper one for the Higgs
 spectrum~{\bf I}, the lower one for the spectrum~{\bf III}),
 and by both production mechanisms (red dashed lines).
}}
\label{figPLOT:QCvsHMED} 
%
\vspace*{0.5truecm}
%
\begin{center} 
\epsfxsize= 12.5cm 
\leavevmode 
\epsfbox{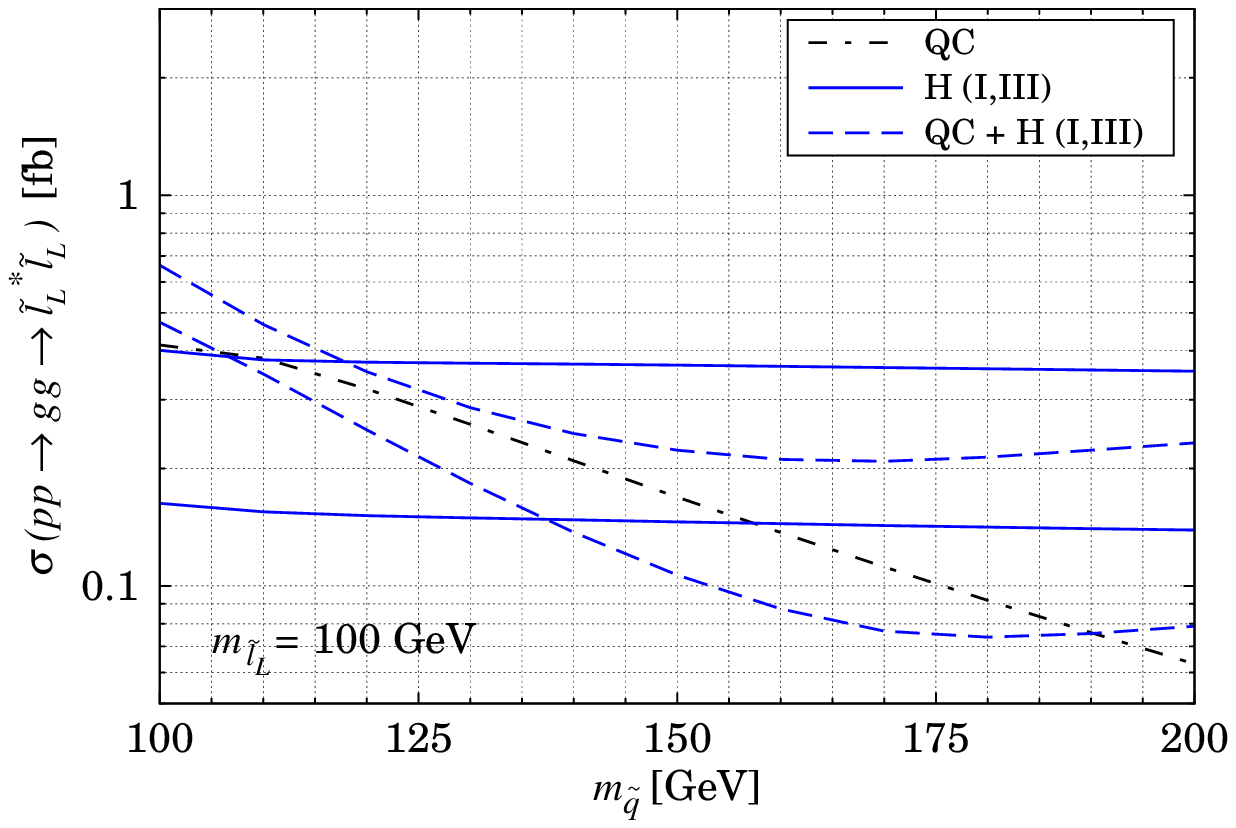} 
\end{center} 
\vspace*{-0.5truecm}
\caption{\small{\it Same as in Fig.~\ref{figPLOT:QCvsHMED} for
 left-handed sleptons.
}}
\label{figPLOT:QCvsHMEDleft} 
\end{figure}

We notice also that, 
 $\vert g_{{\cal H} \widetilde{t}_L \widetilde{t}_L} \vert,
  \vert g_{{\cal H} \widetilde{t}_R \widetilde{t}_R}\vert
 \gsim \vert Y_{{\cal H} tt}\vert$.
Therefore a comparison of the relative size of 
 $A_{S,\,{\cal H}}^{\widetilde{l}_i \widetilde{l}_i}$ and
 $\widetilde{A}_{S,\,{\cal H}}^{\widetilde{l}_i \widetilde{l}_i}$, 
 amounts practically to a comparison of 
 $(v/m_t) F(\hat{s}, m_t^2)$ versus 
 $\sum_i(v^2/\hat{s})S(\hat{s},\widetilde{m}_{t_i}^2)$, 
 $i=1,2$.
The $\widetilde{t}$ squark loop is clearly penalized by a factor
 $(v^2/\hat{s})$ with respect to the factor $(v/m_t)$, constant in
 $\hat{s}$.
Moreover, as Fig.~\ref{fig:SFPfuncs} shows, when integrated in
 $\hat{s}$, the function $S$ gives a smaller  contribution to
 $\widetilde{A}_{S,\,{\cal H}}^{\widetilde{l}_i \widetilde{l}_i}$ than the
 function $F$ gives to
 ${A}_{S,\,{\cal H}}^{\widetilde{l}_i \widetilde{l}_i}$, at least
 as far as the real parts of these functions are concerned. 
In contrast, the imaginary part of $S$ can contribute more than the
 imaginary part of $F$ for very light squarks,
 {\it i.e.} $\sim 100\,$GeV.

Thus, if the squark sector has vanishing chirality-mixing terms, we
 expect the dominant contribution to 
 ${A}_{S}^{\widetilde{l}_i \widetilde{l}_i}$ to be induced by the
 $t$-quark loop, followed by an in-general-smaller contribution coming 
 from the $\widetilde{t}$-squark loop, with exact size depending on
 $\hat{s}_{\rm min}$ and the two $\widetilde{t}$-squark masses.
Finally there is a much smaller contribution from the
 $\widetilde{b}$-squark loop, at least as far as 
 $\widetilde{m}_{b_i} \simeq \widetilde{m}_{t_i}$ for $(i=1,2)$.

Since $(h_E)_{3,3}^2\cos \beta$ is also small,
 the production of same-chirality sleptons
 proceeds always through the D-term contributions to the couplings
 $g_{{\cal H}\widetilde{l}_i \widetilde{l}_i}$, for any of the values of 
 $\tan \beta $ considered here ($\tan \beta \lsim 3$).
Therefore, the size of 
 $\vert\widetilde{A}_{S,\,{\cal H}}^{\,\widetilde{l}_i \widetilde{l}_i}\vert$
 from  $\widetilde{t}$-squark exchange is expected to be in the same
 ballpark of the size of 
 $\vert \widetilde{A}_{\,\rm QC}^{\,\widetilde{l}_i \widetilde{l}_i}\vert$ obtained
 before, except in the resonant region $\hat{s} \sim m_{\cal H}^2$,
 where 
 $\vert\widetilde{A}_{S,\,{\cal H}}^{\,\widetilde{l}_i \widetilde{l}_i}
  \vert$ has the chance to exceed 
 $\vert\widetilde{A}_{\,\rm QC}^{\,\widetilde{l}_i \widetilde{l}_i}\vert$
 substantially.
We have assumed here that also the slepton sector has vanishing
 chirality-mixing terms.

\begin{figure}[t] 
\begin{center} 
\epsfxsize= 12.5cm 
\leavevmode 
\epsfbox{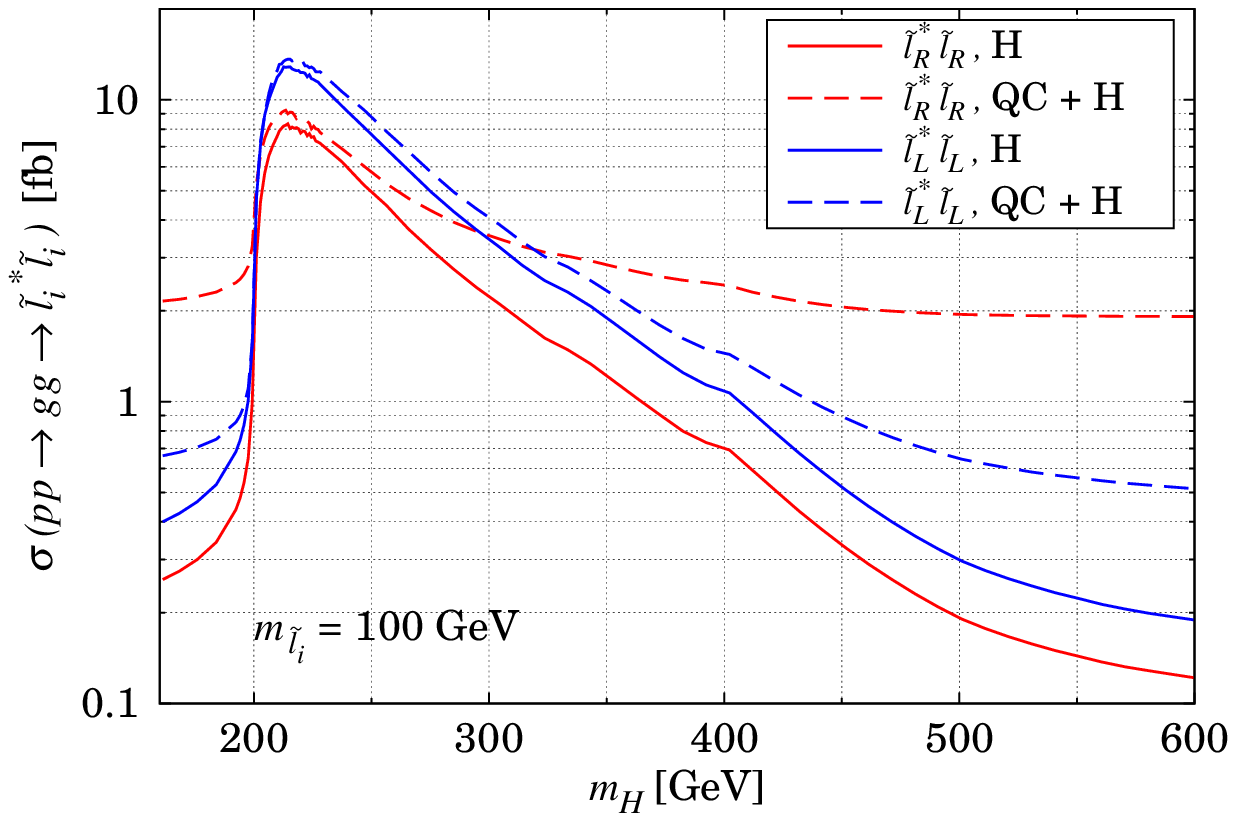} 
\end{center} 
\vspace*{-0.5truecm}
\caption{\small{\it LHC cross sections for the production of
 right- and left-handed sleptons versus $m_H$. 
 The solid lines show the cross sections from Higgs exchange only, the
 dashed lines from both, Higgs exchange and quartic couplings.
 For each of these two groups of curves, those which are lower at the
 peak of the resonance correspond to the production of right-handed
 sleptons, the higher curves, to that of left-handed sleptons.  
 The squark spectrum~{\bf A} with $\widetilde{m}_{q}=100\,$GeV, which 
 induces $\widetilde{m}_{t_1}\simeq 200\,$GeV (see text), was used
 for the evaluation of all squark loops.
}} 
\label{figPLOT:QCHmres} 
\end{figure}

These observations are confirmed by the results shown in
 Figs.~\ref{figPLOT:QCvsHMED} and~\ref{figPLOT:QCvsHMEDleft}, where we
 plot the cross sections for the  production of
 right-handed and left-handed sleptons, respectively, versus
 $\widetilde{m}_q$, which is defined in Sec.~\ref{sec:QCXsecNUMER}.
In both cases we give these results for the Higgs spectrum~{\bf I}
 and~{\bf III}.
For spectrum~{\bf I}, we have used the values
 $m_h =98\,$GeV,  $m_H =162\,$GeV, and $\tan \alpha \sim -1$.
In spectrum~{\bf III}, it is  $m_h =114\,$GeV.

The Higgs-mediated cross sections, shown by the red and blue solid
 lines, are almost horizontal, at $\sim 0.2\,$fb for the Higgs
 spectrum~{\bf I}, and $\sim 0.1\,$fb for the Higgs spectrum~{\bf III}.
That is, in spite of the fact that spectrum~{\bf III} contributes with
 the exchange of $h$ only, whereas spectrum~{\bf I} contributes with
 $h$ and $H$, the value of the Higgs-exchange cross section is very
 similar in the two cases, provided the two scalar Higgs bosons in
 spectrum~{\bf I} are practically degenerate and have mass very similar
 to that of the light Higgs in spectrum~{\bf III}, {\it i.e.}
 $m_h^{\rm I} \sim m_H^{\rm I} \sim m_h^{\rm III}$. 
This is true for both,
 $A_{\rm H}^{\widetilde{l}_i \widetilde{l}_i}$ and
 $\widetilde{A}_{\rm H}^{\widetilde{l}_i \widetilde{l}_i}$.
It is easy to convince oneself of this by inspecting the relevant
 couplings in Tables~\ref{table:GEHiggsSCALcoupls}
 and~\ref{table:MEHiggsSCALcoupls}.
A splitting between $m_h^{\rm I}$ and $m_H^{\rm I}$, however, is in
 general present in spectrum~{\bf I}, with size depending on the
 value of $\tan \beta$ and the squark spectrum.
This accounts for the roughly  $50\%$ difference in the cross sections
 obtained for the Higgs spectrum~{\bf I} and~{\bf III}.
Being practically independent of $\widetilde{m}_{q}$, the Higgs-mediated 
 cross sections show explicitly that the contribution from 
 $\widetilde{A}_{\rm H}^{\widetilde{l}_i \widetilde{l}_i}$
 is small compared to that from 
 ${A}_{\rm H}^{\widetilde{l}_i \widetilde{l}_i}$, as already anticipated.

The quartic-coupling cross sections are shown by the black dot-dashed
 lines. 
They coincide with those in Fig.~\ref{figPLOT:QCxscA}.

The total gluon-initiated cross sections are given by the two
 red (blue) dashed lines, the upper one for spectrum~{\bf I}, the lower
 one for spectrum~{\bf III}.
Although the interference effects can be significant, they are similar
 in size to the corresponding ones induced by quartic couplings only.
We remind that the squark spectrum~{\bf A} used for these figures 
 was chosen in such a way to maximize the size of 
 $\vert\widetilde{A}_{\,\rm QC}^{\,\widetilde{l}_i \widetilde{l}_i}\vert$, 
 but not necessarily that of 
 $\vert\widetilde{A}_{S,\,{\cal H}}^{\widetilde{l}_i \widetilde{l}_i}\vert$.
Indeed, the value of the lightest $\widetilde{t}$-squark in the loop
 ranges between 200--260$\,$GeV when $\widetilde{m}_q$ spans the 
 interval 100--200$\,$GeV.

Enhancements of these cross sections are possible in the region in which
 the Higgs boson $H$ and/or $A$ are resonant.
The cross sections in this region are very sensitive to the widths   
 $\Gamma_H$, which we have evaluated with the program
 CPSUPERH~\cite{CPSUPERH}.

We plot in Fig.~\ref{figPLOT:QCHmres} the cross sections obtained for
 the production of same-chirality sleptons at the lowest point in
 the squark spectrum~{\bf A}, as a function of $m_H$.
The smallest value of $m_H$ is that of the Higgs spectrum~{\bf I} of
 Figs.~\ref{figPLOT:QCvsHMED} and~\ref{figPLOT:QCvsHMEDleft}, the largest 
 value, is quite close to the $m_H$ of spectrum~{\bf III} in these same
 figures. 
No exchange of the pseudoscalar Higgs boson $A$ is possible in this 
 case. 
The production of right-handed sleptons is shown by red lines, that of
 left-handed sleptons by blue lines.
The solid lines show the cross sections induced by Higgs exchange only, 
 the dashed lines show the cross sections due to both production 
 mechanisms. 
The enhancement over the values of the cross sections shown in 
 Figs.~\ref{figPLOT:QCvsHMED} and~\ref{figPLOT:QCvsHMEDleft} is 
 considerable for a rather large range $m_H$.
Although the $\widetilde{t}$-squark loop gives a subdominant
 contribution to the Higgs-exchange cross section with respect to
 $t$-quark loop, a smeared peak is nevertheless visible at
 $m_H \sim 2 \widetilde{m}_{t_1}$.
Far less visible is that at $m_H \sim 2 m_{t}$, because of the smoother
 behaviour at $\hat{s} \sim 2 m_t$ of the real and imaginary part of the
 function $F$, with respect to that in the case of the function $S$.
(See Fig.~\ref{fig:SFPfuncs}.)
At the peak of these resonances, the production cross sections for a 
 pair of right-handed sleptons are roughly twice as large as those
 for a pair of left-handed sleptons. 
This is mainly due to the fact that
 $\vert d^{H_i^0 H_i^0\!,\widetilde{l}_R \widetilde{l}_R}/
        d^{H_i^0 H_i^0\!,\widetilde{l}_L \widetilde{l}_L} \vert = 
 (g_2^2/4 - g_1^2/4)/(g_1^2/2)$, for $(i=u,d)$.

As in the case of those shown in Fig.~\ref{figPLOT:QCxscA}, also these
 cross sections apply to the production of slepton of first and 
 second generation, as well as those of third generation for low
 $\tan \beta$. 
A value $\tan \beta =3$, for example, suppresses the 
 $\tau$-lepton Yukawa couplings sufficiently to reduce $r_\tau$~(see
 discussion in Sec.~\ref{sec:QCXsecNUMER}) to be about few percent, as
 requested in the squark spectrum~{\bf A}.
A moderate cancellation between $(A_L)_{3,3}$ and $\mu \tan \beta$ takes
 place in the chirality mixing terms of the $\widetilde{\tau}$ sector,
 for a value of $\vert \mu \vert \simeq 100\,$ GeV and
 $\vert (A_L)_{3,3}\vert\simeq 200\,$GeV, presumably not too large to
 jeopardize the metastability of the color- and charge-preserving
 vacuum.

The suppression of $(h_E)_{3,3}$, however, is not sufficient to reduce
 the couplings 
 $g_{H\widetilde{\tau}_R \widetilde{\tau}_L}$, 
 $g_{A\widetilde{\tau}_R \widetilde{\tau}_L}$ and their conjugates, in
 which the parameters $(A_L)_{3,3}$ and $\mu$ appear in different
 combinations, 
 to be negligible with respect, for example, to the couplings 
 $g_{H\widetilde{\tau}_R \widetilde{\tau}_R}$, 
 $g_{A\widetilde{\tau}_R \widetilde{\tau}_R}$ coming mainly from D-terms.

Thus, the production of opposite chirality $\widetilde{\tau}$-sleptons
 is possible even for chirality-mixing terms in both, the  
 $\widetilde{\tau}$-slepton mass matrix as well as in
 squark mass matrices that are nearly vanishing, and for a relatively 
 low value of $\tan \beta$. 
This is in contrast to the quartic-coupling case, in which the 
 the production of opposite-chirality $\widetilde{\tau}$ sleptons
 requires a large value of $\tan \beta$.

\begin{figure}[t] 
\begin{center} 
\epsfxsize= 12.5cm 
\leavevmode 
\epsfbox{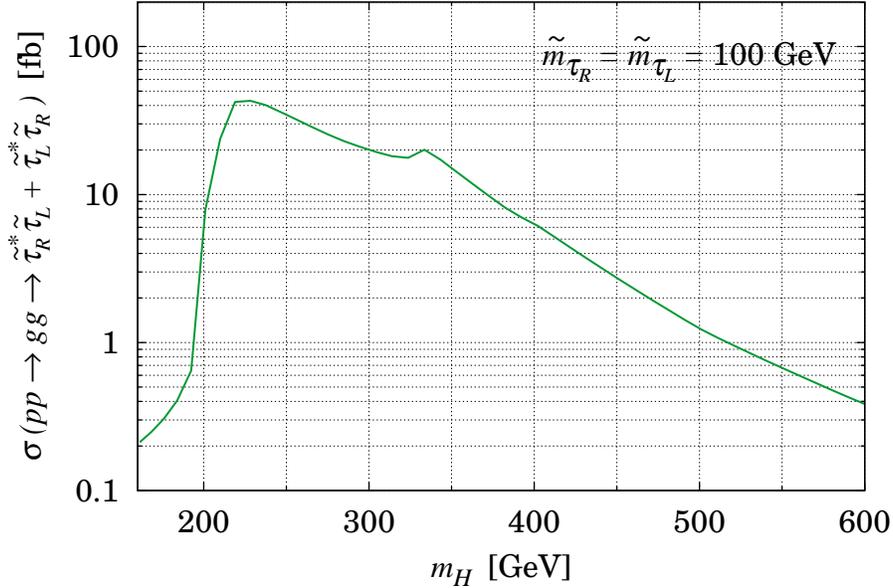} 
\end{center} 
\vspace*{-0.5truecm}
\caption{\small{\it LHC cross section for the production of
 opposite-chirality $\widetilde{\tau}$'s versus $m_H$, obtained at the
 lowest point of the squark spectrum~{\bf A} for $\tan \beta =3$.
 The contribution from quartic couplings is identically vanishing. 
}}
\label{figPLOT:HmOppChir} 
\end{figure}

Indeed, not only is the production of opposite-chirality
 $\widetilde{\tau}$'s possible in this case, but it has a cross section
 larger than that shown for the same spectrum in
 Fig.~\ref{figPLOT:QCHmres}, at least in the resonant Higgs region.
This is because both the two heavier Higgs bosons, $H$ and $A$, can 
 mediate the $t$-quark contribution to the cross section. 
Notice that, because no chirality-mixing terms are present in the 
 squark sector, the exchange of the pseudoscalar Higgs boson is possible
 only with the $t$-quark loop. 
Given the sharp behaviour of the function $P$ shown in
 Fig.~\ref{fig:SFPfuncs}, the peak at the $t$-quark threshold is now
 considerably more pronounced than in the case of the cross section of
 Fig.~\ref{figPLOT:QCHmres} with exchange of only the Higgs boson $H$.

The values of the cross section that this last figure shows on top 
 of the Higgs resonance are now quite interesting. 
We expect these values, as well as those for the production of
 same-chirality sleptons, to increase even further for nonvanishing 
 chirality-mixing terms in the squark sector.
Depending on the values of the trilinear soft parameters, $\mu$, and
 $\tan \beta$, also the $\widetilde{b}$-squark contribution may become 
 competitive with that due to the $\widetilde{t}$-squark and possibly
 also the $t$-quark.

The results shown here are rather promising. 
We believe these cross sections should be studied in full generality
 for more realistic sfermion spectra than the one considered here, and
 for all values of $\tan \beta$, in order to ascertain the possibility
 of probing, through them, important elements of the supersymmetric
 parameter space.

\section{Summary and outlook}
\label{sec:Conclu}
%
We have discussed the possibility of producing slepton pairs through 
 gluon fusion. 
Our guiding motivation was that of assessing the impact that the
 quartic couplings
 $\widetilde{q}^\ast\widetilde{q}\widetilde{l}^\ast\widetilde{l}$ can
 have at the loop level for slepton production. 
If unmistakably detected, coupling such as these could provide an 
 important confirmation of supersymmetric models.

We have found that the corresponding cross sections are small. 
That due to the only sizable quartic coupling from F-terms is small
 because it is proportional to the square of the product of the Yukawa
 couplings of the $b$ quark and the $\tau$ lepton. 
The slepton produced are in this case $\widetilde{\tau}$ sleptons.  
The values of these couplings can be increased for large $\tan\beta$.
Nevertheless, the cross section gets in general penalized by the 
 mixing effects in the slepton and the squark sector, in spite of 
 the fact that this last mixing is necessary to produce the effective
 coupling $gg\widetilde{l}^\ast\widetilde{l}$.

The cross section for slepton production is small also when it is
 induced by quartic couplings from D-terms.
The reason is due to a peculiar cancellation mechanism.
Like the Glashow--Iliopoulos-Maiani (GIM) mechanism cancels the
 contributions from degenerate fermions of up or down-type quarks with
 different flavour in loop-induced flavour-changing-neutral-current
 processes, this mechanism cancels the contributions to
 slepton production from squarks that are degenerate within each
 generation.
The two cancellation mechanisms have however quite different origins. 
The GIM mechanism is due to the unitarity of the 
 Cabibbo-Kobayashi-Maskawa mixing matrix, whereas the cancellation
 observed in slepton production is due to the fact that the sum of the
 SM charges of all quark fields in each family vanishes identically.

In contrast, what we had hoped to be only a supersymmetric background
 to the gluon-fusion production of slepton pairs through quartic
 couplings, {\it i.e.} the gluon-fusion production with the exchange of
 a Higgs boson, turns out to give interesting results. 
The cross section can reach up to ${\cal O}(10)\,$fb if the exchanged
 Higgs boson becomes resonant. 
This value is obtained at the peak of the resonance, for a slepton mass
 of $100\,$GeV, for vanishing chirality-mixing terms in the squark
 sector. 
In this case, the dominant contribution comes from the $t$-quark loop
 that of the $\widetilde{t}$-squark is subdominant, and the 
 $\widetilde{b}$-squark contribution is practically negligible. 
The same spectrum gives a cross section up to $40\,$fb for the 
 production of a pair of opposite-chirality $\widetilde{\tau}$'s even
 for $\tan \beta=3$ and chirality mixing terms in the
 $\widetilde{\tau}$ sector practically negligible. 
Notice with the same type of slepton spectrum, only same-chirality 
 sleptons can be produced through the Drell-Yan production
 mechanism~\cite{DrellYan}.

We have not given explicit values of the cross sections when 
 chirality-mixing terms in the squarks and slepton sectors are
 nonvanishing, and for large values of $\tan \beta$, but we have argued
 why we expect the cross section to increase in this case, at least for
 sleptons still relatively light.
A full-fledged analysis should be performed with more realistic squark
 spectra than those used here, which were devised to maximize the
 gluon-fusion cross sections from quartic-couplings.
Moreover, when $\tan \beta$ is large, the tree-level process
 $b \bar{b} \to $ mediated by Higgs boson exchange, 
 should also be added, incoherently, to the previous processes, in 
 order to give a correct assessment of the values that the 
 gluon-initiated cross section can reach.
This is because the $b$ quark is contained in the proton via a gluon 
 and the $b\bar{b}$-initiated cross section is of the same order in 
 $\alpha_s$ than those studied here.

Finally, where the gluon-fusion mechanism of slepton production stands
 with respect to the Drell--Yan production depends also strongly on 
 the role played by the QCD corrections to the two types of processes. 
Higher-order QCD corrections should indeed be included, together with 
 SUSY-QCD corrections. 
The QCD corrections for the gluon-fusion mechanism are as those for the
 production of Higgs bosons~\cite{HprodNNLO-1,HprodNNLO-2}.
They are known to be large, more than the corrections that QCD induces
 in the case of the Drell--Yan production~\cite{DYNNLO}. 
The SUSY-QCD corrections are also well known~\cite{SUSY-QCD}.  
At large $\tan \beta$ those to the Higgs boson vertex may be well
 approximated~\cite{BGY} by the usual Yukawa coupling corrections.

It is widely believed that, if at all, sleptons will be discovered at
 the LHC through cascade decays. 
The production of a pair of sleptons, if kinematically allowed, 
 will provide their direct mass measurement.
Not much more than this can be learned from the Drell-Yan production.
The gluon-fusion production mechanism, in this respect is potentially
 much more interesting since it can provide valuable information about
 $A$, $\mu$, and $\tan\beta$, and confirmations about the Higgs
 spectrum.
Both mechanisms contribute to the pair production of sleptons, and the
 total cross section may be larger than previously thought, at least
 for resonant Higgs exchange. 
More detailed studies on this subject are clearly needed.

Going back to the issue of quartic couplings, an extended gauge
 structure, for example with an anomalous U(1)$_Y$ in which the
 cancellation mechanism can be avoided, may help with their detection. 
It is however possible that the hadron collider environment is not the
 best suited for this purpose.
The possibilities that a linear collider may offer in this respect
 should be explored.

In the case of its photon-collider option, for example, it is easy to 
 find a situation that parallels very closely the one described here 
 of a pair of sleptons, which, in spite of not being sensitive to
 strong interactions, can nevertheless couple at the one-loop level to
 a pair of gluons, thanks to their quartic couplings to squark pairs. 
The situation that immediately comes to mind is that of a pair of 
 neutral scalars that can be coupled at the one-loop level to two 
 photons thanks to their quartic couplings to a pair of charged 
 scalars.

If the two neutral particles are supersymmetric Higgs bosons, the
 situation can actually be much better than that of slepton production 
 discussed here. 
Indeed, the cancellation mechanism above described, which does work if
 the charged particles exchanged in the loop are squarks, is violated
 when these scalars are sleptons. 
This is simply due to the fact that the neutral components of the 
 SU(2)$_L$ slepton doublets cannot couple to photons.

The production of a pair of supersymmetric Higgs bosons in a photon 
 collider was already considered in several
 papers~\cite{phcollHPP,HPPcomplete}.  
The charged slepton contribution to the loop, is mentioned only in
 Ref.~\cite{HPPcomplete}. 
It is, however, difficult to understand from this analysis the 
 relevance of this contribution. 
It may remain buried under possibly larger
 contributions, not only those due to the exchange of Higgs bosons, 
 but also those coming from box diagrams, that have no correspondent in
 our case of slepton production.
A check of the relative size of all these contributions would probably
 be worthwhile.

\vspace*{1truecm} 
\noindent 
{\large {\bf Acknowledgements}} \\
The authors thank E.~Accomando, B.~Allanach, K.~Fujii, J.S.~Lee,
 K.~Odagiri, N.~Okamura, N.~Polonsky, and K.~Yokoya for
 stimulating discussions and inputs, D.W.~Jung for initial
 collaboration, and M.~Spira for a critical reading of the manuscript.
F.~B. acknowledges the NCTS-KEK Exchange program, which has made
 this collaboration possible and the hospitality extended to her by
 the Yukawa Institute in Kyoto, during the YKIS~2009.
The work of F.B. was partially supported by the Excellency Research
 Project of National Taiwan University, Taiwan:
 ``Mass generation, heavy flavours, neutrinos at the particle physics
 frontier'', grant No~97R0066-60.
K.H. is supported in part by the Grant-in-Aid for scientific
 research~(No. 20340064) from MEXT, Japan. 
 
%

{\small
\begin{thebibliography}{99} 
 
\bibitem{SameSIGNALS}
 H.~Murayama,
{\it Confusing signals of Supersymmetry},
 International Linear Collider Workshop (LCWS2000), 
 Oct. 24-28, 2000, Fermilab,
\\
 http:$//$conferences.fnal.gov/lcws2000/web/P3\_Murayama/index.html.


\bibitem{Review}
  L.~T.~Wang and I.~Yavin,
  {\it A Review of Spin Determination at the LHC},
  Int.\ J.\ Mod.\ Phys.\  A {\bf 23} (2008) 4647


\bibitem{BIBLspinDET}
Papers cited in Ref~\cite{Review} and in:  \\
  F.~Boudjema and R.~K.~Singh,
  {\it A model independent spin analysis of fundamental particles
       using azimuthal asymmetries},
  JHEP {\bf 0907} (2009) 028,
\\
  S.~Y.~Choi, K.~Hagiwara, H.~U.~Martyn, K.~Mawatari and P.~M.~Zerwas,
  {\it Spin analysis of supersymmetric particles},
  Eur.\ Phys.\ J.\  C {\bf 51} (2007) 753



\bibitem{HIGGSmedGGone}
  F.~del Aguila and L.~Ametller,
  {\it On the detectability of sleptons at large hadron colliders},
  Phys.\ Lett.\  B {\bf 261} (1991) 326


\bibitem{HIGGSmedGGtwo}
  M.~Bisset, S.~Raychaudhuri and P.~Roy,
  {\it Higgs-mediated Slepton Pair-production at the Large Hadron
       Collider},
  arXiv:hep-ph/9602430


\bibitem{DrellYan}
  E.~Eichten, I.~Hinchliffe, K.~D.~Lane and C.~Quigg,
  {\it Super Collider Physics},
  Rev.\ Mod.\ Phys.\  {\bf 56} (1984) 579
  [Addendum-ibid.\  {\bf 58} (1986) 1065],
\\
  S.~Dawson, E.~Eichten and C.~Quigg,
  {\it Search For Supersymmetric Particles In Hadron-Hadron Collisions},
  Phys.\ Rev.\  D {\bf 31} (1985) 1581,
\\
  P.~Chiappetta, J.~Soffer and P.~Taxil,
  {\it Spin Asymmetries For Scalar Leptons From W And Z Decay In P Anti-P
      Collisions},
  Phys.\ Lett.\  B {\bf 162} (1985) 192,
\\
  H.~Baer, C.~h.~Chen, F.~Paige and X.~Tata,
  {\it Detecting sleptons at hadron colliders and supercolliders},
  Phys.\ Rev.\  D {\bf 49} (1994) 3283


\bibitem{CHH}
  M.~Claudson, L.~J.~Hall and I.~Hinchliffe,
  {\it Low-Energy Supergravity: False Vacua And Vacuous Predictions},
  Nucl.\ Phys.\  B {\bf 228} (1983) 501,
\\
  J.~A.~Casas, A.~Lleyda and C.~Munoz,
  {\it Strong constraints on the parameter space of the MSSM from charge
       and color breaking minima},
  Nucl.\ Phys.\  B {\bf 471} (1996) 3


\bibitem{LargeAtunn}
  A.~Kusenko, P.~Langacker and G.~Segre,
  {\it Phase Transitions and Vacuum Tunneling Into Charge and Color
       Breaking Minima in the MSSM},
  Phys.\ Rev.\  D {\bf 54} (1996) 5824,
\\
  U.~Sarid,
  {\it Tools for tunneling},
  Phys.\ Rev.\  D {\bf 58} (1998) 085017
  [arXiv:hep-ph/9804308],  
\\
  F.~Borzumati, G.~R.~Farrar, N.~Polonsky and S.~D.~Thomas,
  {\it Soft Yukawa couplings in supersymmetric theories},
  Nucl.\ Phys.\  B {\bf 555} (1999) 53


\bibitem{HIKASA}
  K.~Hikasa, 
 {\it Supersymmetric Standard Model for Collider Physicists}, 
 to be requested to the author,
\\
  S.~P.~Martin,
  {\it A Supersymmetry Primer},
  arXiv:hep-ph/9709356



\bibitem{CTEQ}
  J.~Pumplin, D.~R.~Stump, J.~Huston, H.~L.~Lai, P.~Nadolsky and
  W.~K.~Tung,
  {\it New generation of parton distributions with uncertainties from
       global QCD analysis},
  JHEP {\bf 0207} (2002) 012


\bibitem{BASES}
  S.~Kawabata,
  {\it A New Monte Carlo Event Generator For High-Energy Physics},
  Comput.\ Phys.\ Commun.\  {\bf 41} (1986) 127 
and
  {\it A New version of the multidimensional integration and event
    generation package BASES/SPRING},
  Comput.\ Phys.\ Commun.\  {\bf 88} (1995) 309


\bibitem{SPIRAhabil}
  M.~Spira,
  {\it QCD effects in Higgs physics},
  Fortsch.\ Phys.\  {\bf 46} (1998) 203


\bibitem{ABDELD}
 A.~Djouadi,
 {\it The anatomy of electro-weak symmetry breaking. II:
      The Higgs bosons in the minimal supersymmetric model},
  Phys.\ Rept.\  {\bf 459} (2008) 1,
\\
  M.~S.~Carena {\it et al.}  [Higgs Working Group Collaboration],
  {\it Report of the Tevatron Higgs working group},
  arXiv:hep-ph/0010338


\bibitem{CPSUPERH}
  J.~S.~Lee, A.~Pilaftsis, M.~S.~Carena, S.~Y.~Choi, M.~Drees,
  J.~R.~Ellis and C.~E.~M.~Wagner,
  {\it CPsuperH: A computational tool for Higgs phenomenology in the
       minimal supersymmetric standard model with explicit CP violation},
  Comput.\ Phys.\ Commun.\  {\bf 156} (2004) 283


\bibitem{HprodNNLO-1}
  D.~Graudenz, M.~Spira and P.~M.~Zerwas,
  {\it QCD corrections to Higgs boson production at
       proton proton colliders},
  Phys.\ Rev.\ Lett.\  {\bf 70} (1993) 1372,
\\
  M.~Spira, A.~Djouadi, D.~Graudenz and P.~M.~Zerwas,
  {\it Higgs boson production at the LHC},
  Nucl.\ Phys.\  B {\bf 453} (1995) 17,
\\
  R.~V.~Harlander and K.~J.~Ozeren,
  {\it Top mass effects in Higgs production at next-to-next-to-leading
      order QCD: virtual corrections},
  Phys.\ Lett.\  B {\bf 679} (2009) 467,
 \\
 A.~Pak, M.~Rogal and M.~Steinhauser,
  {\it Virtual three-loop corrections to Higgs boson production
       in gluon fusion for finite top quark mass},
  Phys.\ Lett.\  B {\bf 679} (2009) 473,
\\
  S.~Catani, D.~de Florian, M.~Grazzini and P.~Nason,
  {\it Soft-gluon resummation for Higgs boson production at
      hadron colliders},
  JHEP {\bf 0307} (2003) 028


\bibitem{HprodNNLO-2}
  S.~Dawson, A.~Djouadi and M.~Spira,
  {\it QCD corrections to SUSY Higgs production:
            The Role of squark loops},
  Phys.\ Rev.\ Lett.\  {\bf 77} (1996) 16,
\\
  M.~Muhlleitner and M.~Spira,
  {\it Higgs boson production via gluon fusion: Squark loops at NLO QCD},
  Nucl.\ Phys.\  B {\bf 790} (2008) 1,
\\
  R.~Bonciani, G.~Degrassi and A.~Vicini,
  {\it Scalar Particle Contribution to Higgs Production via
        Gluon Fusion at NLO},
  JHEP {\bf 0711} (2007) 095,
\\
  M.~Muhlleitner, H.~Rzehak and M.~Spira,
  {\it MSSM Higgs Boson Production via Gluon Fusion:
        The Large Gluino Mass Limit},
  JHEP {\bf 0904} (2009) 023


\bibitem{DYNNLO}
  W.~Beenakker, M.~Klasen, M.~Kramer, T.~Plehn, M.~Spira and P.~M.~Zerwas,
  {\it The Production of charginos / neutralinos and sleptons at hadron
        colliders},
  Phys.\ Rev.\ Lett.\  {\bf 83} (1999) 3780
  [Erratum-ibid.\  {\bf 100} (2008) 029901],
\\
  M.~Grazzini,
  {\it The Drell-Yan process in NNLO QCD},
  arXiv:0908.1336 [hep-ph]


\bibitem{SUSY-QCD}
  R.~V.~Harlander and M.~Steinhauser,
  {\it Supersymmetric Higgs production in gluon fusion at
       next-to-leading order},
  JHEP {\bf 0409} (2004) 066,
\\
  R.~V.~Harlander and F.~Hofmann,
  {\it Pseudo-scalar Higgs production at next-to-leading order
         SUSY-QCD},
  JHEP {\bf 0603} (2006) 050,
\\
 G.~Degrassi and P.~Slavich,
  {\it On the radiative corrections to the neutral Higgs boson
       masses in the NMSSM},
  Nucl.\ Phys.\  B {\bf 825} (2010) 119,
\\
  C.~Anastasiou, S.~Beerli and A.~Daleo,
  {\it The two-loop QCD amplitude gg $\to$ h,H in the Minimal
       Supersymmetric Standard Model},
  Phys.\ Rev.\ Lett.\  {\bf 100} (2008) 241806



\bibitem{BGY}
  M.~S.~Carena, D.~Garcia, U.~Nierste and C.~E.~M.~Wagner,
  {\it Effective Lagrangian for the $\bar{t} b H^{+}$ interaction
    in the MSSM and charged Higgs phenomenology},
  Nucl.\ Phys.\  B {\bf 577} (2000) 88,
\\
 J.~Guasch, P.~Hafliger and M.~Spira,
  {\it MSSM Higgs decays to bottom quark pairs revisited},
  Phys.\ Rev.\  D {\bf 68} (2003) 115001,
\\
  F.~Borzumati, C.~Greub and Y.~Yamada,
  {\it Beyond leading-order corrections to $\bar{B} \to X_s \gamma$ at
      large tan(beta): The charged-Higgs contribution},
  Phys.\ Rev.\  D {\bf 69} (2004) 055005,
and
  {\it Towards an exact evaluation of the supersymmetric
       O(alpha(s) tan(beta)) corrections to $\bar{B}\to X_s \gamma$},
  arXiv:hep-ph/0305063
\\
  D.~Noth and M.~Spira,
  {\it Higgs Boson Couplings to Bottom Quarks: Two-Loop Supersymmetry-QCD
        Corrections},
  Phys.\ Rev.\ Lett.\  {\bf 101} (2008) 181801

 
\bibitem{phcollHPP}
  S.~H.~Zhu,
  {\it Pseudoscalar Higgs boson pair production in photon photon
       collisions},
  J.\ Phys.\ G {\bf 24} (1998) 1703,
\\
  S.~H.~Zhu, C.~S.~Li and C.~S.~Gao,
  {\it Lightest neutral Higgs pair production in photon photon collisions
     in the minimal supersymmetric standard model},
  Phys.\ Rev.\  D {\bf 58} (1998) 015006,
\\
  G.~J.~Gounaris and P.~I.~Porfyriadis,
  {\it The gamma gamma $\to$ A0 A0 process at a gamma gamma collider},
  Eur.\ Phys.\ J.\  C {\bf 18} (2000) 181


\bibitem{HPPcomplete} 
  Y.~J.~Zhou, W.~G.~Ma, H.~S.~Hou, R.~Y.~Zhang, P.~J.~Zhou and Y.~B.~Sun,
  {\it Neutral Higgs boson pair production via gamma gamma collision in
    the minimal supersymmetric standard model at linear colliders},
  Phys.\ Rev.\  D {\bf 68} (2003) 093004


\end{thebibliography} 
}
\end{document}